\documentclass[aps,prd,a4paper,onecolumn,amsmath,showpacs,superscriptaddress,nofootinbib,preprintnumbers,notitlepage]{revtex4-1}
\usepackage{amssymb,amsmath}
\usepackage{graphicx}
\usepackage{color}
\usepackage{dcolumn}
\usepackage{pgfplots}
\usepackage{booktabs}
\usepackage{multirow}
\usepackage{dcolumn}
\usepackage{amsmath}
\usepackage{mathtools}
\usepackage{amsfonts}
\usepackage{amssymb}
\usepackage{epstopdf}
\usepackage{bm}
\usepackage{siunitx}
\usepackage{braket}
\usepackage{enumitem}
\usepackage{soul}
\usepackage{ulem}
\usepackage{color}
\usepackage{transparent}
\usepackage{pifont}
\usepackage[font={small}]{caption}
\usepackage{subcaption}
\usepackage{float}

\begin{document}
\title{Speed of sound and scalar spectral index: 
Reconstructing inflation and reheating in  a non-canonical theory}
\author{Ramón Herrera}
\email{ramon.herrera@pucv.cl}
\affiliation{Instituto de F\'{\i}sica, Pontificia Universidad Cat\'{o}lica de Valpara\'{\i}so, Avenida Brasil 2950, Casilla 4059, Valpara\'{\i}so, Chile.
}

\author{Carlos  Ríos}
\email{carlos.rios@ucn.cl}
\affiliation{Instituto de F\'{\i}sica, Pontificia Universidad Cat\'{o}lica de Valpara\'{\i}so, Avenida Brasil 2950, Casilla 4059, Valpara\'{\i}so, Chile.
}
\affiliation{Departamento de Ense\~nanza de las Ciencias B\'asicas, Universidad Cat\'olica del Norte, Larrondo 1281, Coquimbo, Chile.}

\begin{abstract}
In this article we analyze the reconstruction of inflation in the framework of a non-canonical theory. In this sense, we study the viability of reconstructing the background variables assuming a non-linear kinetic term given by $K(X,\phi)=X+g(\phi)X^2$, with $X$ the standard kinetic term associated to the scalar field $\phi$ and $g(\phi)$ an arbitrary coupling  function.
In order to achieve this reconstruction in the context of  inflation, we assume the slow-roll approximation together with the parametrization of the scalar spectral index $n_s$ and the speed of sound $c_s$ as a function of the number of $e-$folds $N$. By assuming the simplest parametrizations for $n_s-1=-2/N$ and $c_s\propto N^{-\beta}$ with  $\beta$ a constant, we find the reconstruction of  the effective potential $V(\phi)$ and the coupling function $g(\phi)$ in terms of the scalar field. Besides, we study the  reheating  epoch by considering  a constant equation of state parameter, where we determine the temperature and number of $e-$folds during the reheating epoch in terms of the reconstructed variables and the observational parameters. In this way, the parameter-space related to the reconstructed  inflationary model are constrained during the epochs of inflation and reheating by assuming the current astronomical data from Planck and BICEP/Keck results.  
\end{abstract}

\maketitle
\section{Introduction}

It is well known that the dynamics and evolution of the early epoch of the  universe can be specified  by the  hot big bang model, nevertheless, this  model posses different cosmological problems (flatness, horizon, monopole, etc.) that the inflationary scenario  resolves assuming  an accelerated expansion before to the radiation era \cite{1A,2a,2b,3A}. Here, the great importance of introducing the inflationary epoch is that this stage gives an account of the large-scale structure, as well  it delivers  a causal description of the anisotropies observed in the cosmic microwave background (CMB) radiation \cite{5,6,7}.

In order to describe the inflationary epoch, we have different models that give an account of an accelerated expansion during  the early universe. In this sense, we can mention those  inflationary models in which the inflationary phase is driven for a canonical or also a non-canonical scalar field \cite{2a,8,9}. In this respect, we can distinguish the k-essence inflation, in which the  action (or Lagrangian density) associated to k-essence  introduces  a nonstandard higher order kinetic term or non-canonical term related with the scalar field  or inflaton field, see Refs.\cite{10,11}. In this respect, a non-linear kinetic term characterized for an arbitrary function  $K(X,\phi)$ which is present in the Lagrangian density  $\mathcal{L}(X,\phi)=K(X,\phi)-V(\phi)$, where $X$ corresponds to the canonical kinetic term associated to scalar field $\phi$ and $V(\phi)$ is the effective potential. In particular the simplest non-linear kinetic term is given by $K(X)=k_{n+1}X^n$ \cite{Mukhanov:2005bu,Li:2012vta}, where $n$ corresponds to an integer number and  the parameter $k_{n+1}$ is a constant such that the quantity $k_{n+1}X^n$ has units of $M_p^4$, with $M_p$ the Planck mass. The situation in which the non-linear kinetic term is given by $K(X,\phi)=K(\phi)X+X^2$, with $K(\phi)$ an arbitrary function of the inflaton field  was studied in the framework of warm inflation in Ref.\cite{Peng:2016yvb}.
Besides, in order to study the k-inflation  different effective potentials related with the scalar field have been analyzed in the literature in the framework of the slow-roll dynamics \cite{SL1}.  It is well known that 
in the framework of a non-canonical theory  a non lineal kinetic term  introduces a reduced speed of sound $c_s$ smaller than the speed of light ($c_s<1$). This reduction on $c_s$ generates a suppression of the tensor to scalar ratio \cite{11} as well as a large amount  of non-Gaussianities during inflation, see Refs.
\cite{Chen:2006nt,DeFelice:2011uc}. 
Thus, any non-canonical theory has an evolving speed of sound, being the simplest the case in which this speed is a constant, see e.g., Refs.\cite{Sergijenko:2014pwa,Kunz:2015oqa}. In  this sense,   the sound speed should  always be subluminal and also a real quantity, since we do not expect that whichever  physical information to propagate faster than the speed of light in a vacuum, with which $0<c_s^2\leq 1$. 
In particular, for the simplest non-linear kinetic term  $K(X)=k_{n+1}X^n$, the speed of sound is given by $c_s^2=\frac{1}{2n-1}=$ constant and it is reduced if the power $n$ becomes  $n>1$, see  e.g.,\cite{Pareek:2021lxz,Herrera:2023mas}.
In relation to the speed of sound, we can find in the literature some expressions, in order to describe the early universe. In particular in Ref.\cite{Shi:2021tmn} the authors utilize the ansatz for the speed of sound given by $c_s\propto (-\tau)^s$, in which $\tau$ corresponds to the conformal time and the power $s$ a constant, to study the duality of cosmological models during inflation. The situation in which the speed of sound becomes $c_s^2\propto\rho^{1-2n}$, with $n$ a constant  associated to the non lineal kinetic  term into the Lagrangian was developed in \cite{Magueijo:2008pm} and the specific case in which $n=-1/2$ the speed  $c_s$ is proportional to the energy density $\rho$. In Ref.\cite{Cai:2009hw} the authors analized different sound speeds $c_{s_I}\propto H/t$ in a model of multiple Dirac-Born-Infeld (DBI) type actions with multiple scalar fields.  For another interesting  variations  of the speed of sound associated to the early universe, see e.g., \cite{Tolley:2009fg,Canas-Herrera:2020mme,Achucarro:2010da}. 
In this form,  the k-essence models allow  
that the speed of sound associated to a non lineal kinetic term is smaller than speed of light or equal to speed of light, see e.g.\cite{10,Deffayet:2011gz,Babichev:2007dw}. 
 In this context, in the literature these models  become  attractive to describe the inflationary epoch as well the present era  characterized by the dark energy \cite{Chimento:2003ta,Melchiorri:2002ux}.

Additionally, in relation to the k-models an important consideration to take into account 
 is related  with the speed gravitational waves and it 
 is equal to the speed of light. Thus, the speed gravitational waves in the k-models   coincides with the speed found  from the detection made by  GW170817 and the $\gamma$-ray burst\cite{GW}. 

In relation to the early universe and in particular to the inflationary epoch, 
the reconstruction of the background quantities, such as the effective potential and  the coupling functions using  the observational variables, such as the scalar spectrum, the scalar spectral index, and the tensor to scalar ratio, have been studied by different authors in the modern cosmology\cite{Au1,Au2,Chiba}.

In this respect, we can mention that  the procedure 
of  reconstruction of the inflation, through  the parametrization of the observational variables  as a function of the number of $e-$folds $N$   is developed  under the slow-roll approximation. Under this formalism, using 
 the parametrization of the observational parameters, we  can say that the scalar spectral index $n_s$ in terms of the number of $e-$ folds $N$ i.e., $n_s(N)$ is the most utilized. Thus, the simplest parametrization for the spectral index $n_s(N)$ corresponds to   $n_s(N)=1-2/N$ for large-$N$ and this index is well confirmed by Planck results, in the case in which the number $N\simeq 55-70$. Here large $N$ corresponds to values of the number $N\sim\mathcal{O}(10)\sim\mathcal{O}(10^2)$ generated 
during the slow roll stage \cite{Chiba}. In this respect, we assume that the number of $e-$folds $N\simeq 55 -70$ denotes to the comoving scale that crossed the Hubble radius during the inflationary era. In the frame of the general relativity (GR), the parametrization of  the scalar spectral index $n_s(N)=1-2/N$ gives rise to the reconstruction of the different  effective potentials associated to the scalar field. 
In this sense, in the frame of GR the parametrization or attractor characterized by the spectral index $n_s(N)=1-2/N$ gives rise to the reconstruction of the effective potential as a function of the scalar field such as; T-model, the E-model, $R^2-$model, the chaotic model, etc. see Refs.\cite{Chiba,M1,M2}. For the reconstruction of two variables associated to the background inflation as in the case of the reconstruction of warm inflation, it was required to utilize two parameters; the scalar spectral index $n_s(N)$ and the tensor to scalar ratio $r(N)$ \cite{Herrera:2018cgi}, see also the reconstruction (two background variables) of $G-$inflation   in Ref.\cite{Herrera:2018mvo}.

However, there are other parametrizations in terms of the number of $e-$folds  $N$ used in the literature, in order to rebuild the background variables associated to the inflationary epoch in the context of slow-roll approximation. For example, we mention the parametrization utilized on the slow roll parameter $\epsilon(N)$, as a function of the number of $e-$ folds $N$ \cite{Au2}, see also \cite{Es2,Es2a,Es3}. Besides, in Ref.\cite{Roe} was studied  the reconstruction of the effective potential and the scalar spectral index considering the parametrization the two slow-roll parameters; $\epsilon(N)$ and the another parameter $\eta(N)$. Similarly, the reconstruction of the effective potential $V(\phi)$ (with $\phi$ the scalar field), utilizing the speed of the scalar field in terms of $N$ as ansatz was analyzed in Ref.\cite{Seba}. For a review of another reconstructions using different quantities,  see Refs.\cite{Es4,Es8,Es9,Gonzalez-Espinoza:2021qnv,Herrera:2022kes, OTROS1,OTROS2,OTROS3}. 

On the other hand, at the end of the inflationary era we have a phase of reheating of the universe, in order to connect with  the standard big-bang model, see \cite{St1}. In this scenario of reheating of the universe, the matter and the radiation are produced via the decay of the inflaton field or other fields, whereas the temperature of the universe during this era increases in magnitude and then the universe enters to the radiation dominated era.  In order to describe the reheating of the universe, there are different mechanisms or reheating models to grow the temperature during  the early universe. In this sense, we have the mechanism related to the perturbative decay of the scalar field from an oscillating process  at the end of inflation    
in which particles are produced\cite{R1}. Also, we have a mechanism associated to a non-perturbative analysis  as the model of instant reheating\cite{R4} or as parametric resonance decay of the inflaton or another field\cite{R2}. Besides, we have some models of inflation  in which the inflaton field does not oscillate around of the minimum of the potential and these   models  are called non-oscillating models (NO models) and the process of reheating takes place from the decay of  another field called curvaton \cite{R5,yo2}, see  also Ref.\cite{R3} for the reheating from tachyonic instability and another models in Ref.\cite{OTROS5}.  

In relation to the reheating epoch, we have some important parameters that characterize this scenario such as; the reheating temperature $T_\text{reh}$, an equation of state (EoS) denoted as $\omega_\text{reh}=p/\rho$ related to the matter content during the reheating era and the  duration of this stage which is distinguished by the number of $e-$folds occurred  during this scenario $N_\text{reh}$. Here, the quantities  $p$ and $\rho$ denote the pressure and energy density of the fluid associate to the matter.

In the case of the reheating temperature, we have a lower bound imposed by primordial nucleosynthesis (BBN), where the temperature in this period   $T_\text{BBN}\sim$ 10 MeV, see Ref.\cite{MeV}. In relation to the EoS parameter $\omega_\text{reh}$, several numerical analyzes were studied for the matter content during the reheating epoch,  in which  different interactions are realized  between the scalar field and other fields during the reheating. In general, we have that this EoS parameter is a function of the time for the different process in the reheating era. Thus, in particular for the situation in which we have a power law potential, the EoS parameter at the end of the inflationary stage becomes $\omega_\text{reh}=0$ and then increases to positive values where the parameter achieves the value  $\omega_\text{reh}\sim 0.3$, see Ref.\cite{Felder:2000hq}. In the situation in which we have a massive field, the authors in Ref.\cite{Felder:2000hq} found that the EoS parameter takes a negative value at the end of inflationary stage   $\omega_\text{reh}=-1/3$ and then increase to the value $\omega_\text{reh}=0$ for a later time, see also \cite{Dodelson:2003vq,Munoz:2014eqa}. As a first approximation and  following Ref.\cite{Munoz:2014eqa}, we will assume  that the EoS parameter can be assumed 
independent of time i.e.,  a constant parameter during this period, 
in order to find analytical expressions for the temperature and $e-$folds during the reheating.

The main goal of the present research is to apply the reconstruction methodology under the slow-roll approximation in the early universe, considering  as  attractor the parametrization of the   the scalar spectral index $n_s(N)$  together with the speed of sound $c_s(N)$, in terms of the number of $e-$folds $N$. In this context, from the framework of a non-canonical theory, we will study  a non-linear kinetic term defined as  $K(X,\phi)=X+g(\phi)X^2$, to rebuild the potential associated to the inflaton field $V(\phi)$ and the coupling function $g(\phi)$. In this way, in the context of  this theory, we will study how the reconstruction of the background variables are modified assuming the parametrizations on the spectral index  $n_s(N)$ and the speed $c_s(N)$, respectively. In this sense, considering a general procedure, we will rebuild the effective potential  $V(\phi)$ and the coupling 
function $g(\phi)$ giving $n_s(N)$ and the speed of sound $c_s(N)$. 

Besides, we will analyze the reheating epoch 
and how the parameters during this stage  are modified from the reconstruction  (background variables) found during the slow-roll inflation. In relation to these reheating parameters, we will determine  the reheating temperature and the duration of the reheating characterized through of the number of $e-$folds during the reheating   and how considering the observational parameters from Planck measurements  these reheating quantities are constrained.

We organize our article as follows: In Section \ref{inf} we give a brief description of the inflationary phase in a non-canonical theory.
Here we show the background equations in the framework of the slow-roll approximation together with the cosmological perturbations in this theory. In Section \ref{reconstruction} we find under a general formalism, expressions for the effective potential and the coupling function associated to the non-linear kinetic term as a function of the number of $e-$folds $N$ to obtain the reconstruction from any parametrization related to the scalar spectral index $n_s(N)$ as well as  the speed of sound $c_s(N)$. In Section \ref{Rec2}, we consider a specific example for both the scalar spectral index $n_s=n_s(N)$ given by $n_s=1-2/N$  and the speed of sound $c_s(N)\propto N^{-\beta}$ with $\beta$ a constant parameter, in order to reconstruct both the effective potential $V(\phi)$ as well as the coupling function associated to the non-linear kinetic term $g(\phi)$. Besides, we obtain the constrains on the space-parameter for the different quantities or parameters associated to our model considering the present astronomical data from Planck and BICEP/Keck results.  
In Section \ref{Reheating}, we analyze the reheating stage considering the background variables found in the previous section. In this section, we find the reheating temperature together with the number of $e-$folds during the reheating epoch. Finally, in Section \ref{Conc} we give our conclusions and remarks. We chose units in which  $c=\hbar=1$.

\section{The inflationary phase in a non-canonical theory}\label{inf}

In this section we will give a brief description  of theory of gravitation described by a non-canonical kinetic term in the framework of the GR. In this context, we start with 
the four dimension action $S$ given by \cite{K1,K2}
\begin{equation}
\label{action}
	S=\int\left[\frac{R}{2\kappa}+\mathcal{L}(X,\phi) \right]\sqrt{-g}\,d^4\,x,
\end{equation}
in which $\kappa=8\pi G=M_p^{-2}$, where recalled that  $M_p$ is the Planck mass,  $R$ denotes the scalar Ricci and $g$ is the determinant of the metric $g_{\mu\nu}$. Besides, the quantity 
$\mathcal{L}(X,\phi)$ corresponds to the Lagrangian density related with the scalar field  $\phi$ and the kinetic energy of the scalar field $X$ defined as $X=(g^{\alpha\beta}\partial_\alpha\phi\partial_\beta\phi)/2$.

From the effective action given by Eq.(\ref{action}), we can identify that the energy density $\rho$ and the pressure $p$ 
as a function of the scalar field $\phi$ and $X$ assuming a perfect fluid for the matter become\cite{K1,K2}
\begin{equation}
\rho=2X\frac{\partial \mathcal{L}(X,\phi) }{\partial X}-\mathcal{L}(X,\phi),\,\,\,\,\,\,\mbox{and}\,\,\,\,\,\,\,\,\,p=\mathcal{L}(X,\phi),\label{rp}
\end{equation}
 respectively. Note that for the specific case in which $\mathcal{L}(X,\phi)=X-V(\phi)$,   with $V(\phi)$  the effective potential, these quantities are reduced to the standard expressions for the energy density $\rho$ and the pressure $p$ associated to a scalar field in the framework of a canonical theory.

In relation to the  Lagrangian density associated to the scalar field this Lagrangian can be  written as an expansion of the form \cite{K1}
\begin{equation}
    \mathcal{L}(X,\phi)=\sum_{n\geq 0}g_n(\phi)X^{n+1}-V(\phi),
    \label{exp1}
\end{equation}
where the quantities $g_n(\phi)$ correspond to  arbitrary functions of the scalar field $\phi$. As before, for the case in which we take $n=0$ and $g_0(\phi)=1$, it is reduced to a canonical theory.

In the following  we will consider the specific values of $n=0$ and $n=1$, where the functions $g_n(\phi)$ are defined as
 $g_0(\phi)=1$ and $g_1(\phi)=g(\phi)$, respectively. Here, just for  simplicity and in order to obtain analytical expressions for the reconstruction of the  background variables, 
 we choose these two terms in the expansion given by Eq.(\ref{exp1}).  In this way,  the Lagrangian density associated to the inflaton field takes the  structure analyzed  in Ref.\cite{K1} (see also Refs.\cite{KK1,KK2})
\begin{equation}
\label{p}
   \mathcal{L}(X,\phi)=K(X,\phi)-V(\phi)=X+g(\phi)X^2-V(\phi),
\end{equation}
where $K(X,\phi)=X+g(\phi)X^2$ corresponds to an arbitrary function associated to a non-linear kinetic term  and the function $g(\phi)$ has units of $\kappa^2=M_p^{-4}$. 
The motivation to consider the Lagrangian density (or action) defined by Eq.(\ref{p}) comes from considering string-loop corrections which generate a non-trivial moduli field dependence of the coefficients of the various kinetic terms. These terms appear due to the massive modes of the string in the low-energy effective action, see Refs.\cite{Damour:1994zq,Damour:1995pd,Foffa:1999dv}. In this form, motivated by these low-energy effective actions, we shall consider as simplest model to rebuild the inflationary epoch in which the Lagrangian density takes the structure of Eq.(\ref{p}).

In this way, from Eq.(\ref{p}), we have that the energy density and pressure are defined as 
\begin{equation}
\rho=X+3g(\phi)X^2+V(\phi),\,\,\,\,\,\mbox{and}\,\,\,\,\,\,\,
p=X+g(\phi)X^2-V(\phi),
\label{p1}
\end{equation}
respectively.

 In order to study the early universe and its dynamics, we consider a spatially flat Friedmann Robertson Walker metric, along  with a homogeneous scalar field i.e., $\phi=\phi(t)$.  Thus, 
 using the Friedmann equation given by $H^2=(\kappa/3) \rho$, where $H=\dot{a}/a $ represents the Hubble parameter and $a=a(t)$ denotes the scale factor, we have
\begin{equation}
    H^2=\frac{\kappa}{3}\left[\frac{1}{2}\dot{\phi}^2+\frac{3}{4}g\dot{\phi}^4
+V\right].
\end{equation}
In the following, the dots represent differentiation with respect to the time $t$ and $g(\phi)=g$.

Additionally, from the continuity equation defined as $\dot{\rho}+3H(\rho+p)=0$ and considering Eqs.(\ref{rp}) and (\ref{p}), this continuity  equation can be rewritten as
\begin{equation}
\label{dotX}
    \dot{X}=\frac{\sqrt{2X}c_s^2}{p_X}\left(2Xp_{X\phi}-p_\phi-3H\sqrt{2X}p_X \right),
\end{equation}
or equivalently
\begin{equation}
\label{ddotPhi}
    \ddot{\phi}+\frac{3H\dot{\phi}(1+g\dot{\phi}^2)}{1+3g\dot{\phi}^2}-\left[\frac{3/4g_\phi\dot{\phi}^4-V_\phi}{1+3g\dot{\phi}^2}\right]=0.
\end{equation}

Here the quantity $c_s^2$ is defined as 
\begin{equation}
\label{cs}
    c_s^2\equiv\frac{p_X}{\rho_X}=\left(1+\frac{2Xp_{XX}}{p_X}\right)=\frac{1+g\dot\phi^2}{1+3g\dot\phi^2},
\end{equation}
and it corresponds to the adiabatic speed of sound squared and this speed depends excursively of the Lagrangian density associated to scalar field. In our model, we note that the speed of sound depends on the coupling function $g$ associated to the non lineal kinetic term and the speed of scalar field $\dot{\phi}$.
Also, 
note that in the limit in which $g\rightarrow 0 $, this speed is reduced to the standard canonical field theory in which $c_s\rightarrow 1$ (speed of light).
In the following, we will utilize that the notation $p_X=\partial p/\partial X$, $p_\phi$
denotes $p_\phi=\partial p/\partial \phi$, $p_{XX}$ to $p_{XX}=\partial^2 p/\partial X^2$, etc.

From Eq.(\ref{cs}), we find that the coupling function $g$ can be  expressed in term of the speed of sound  and the speed of the scalar field as
\begin{equation}
g=\left(\frac{1-c_s^2}{3c_s^2-1}\right)\frac{1}{\dot{\phi}^2},\,\,\,\,\,\mbox{with}\,\,\,\,\,c_s^2\neq\frac{1}{3}.\label{gf}
\end{equation}
Here we note that  from the specific Lagrangian density defined by Eq.(\ref{p}), the above equation shows that the speed of sound  presents an upper and a lower limit given by $1\geqslant c_s^2>1/3$, in order to have a continuous coupling function $g(\phi)$ for all values of the field $\phi$. As before, we note that in the specific value in which  $c_s\rightarrow 1$, it reduces to the standard canonical field theory in which the function $g\rightarrow 0$.

In relation to the expansion of the universe during the inflationary era, we can introduce the number of $e-$folds defined as 
\begin{equation}
\label{deltaN}
    \Delta N=N-N_f=\ln[a(t_f)/a(t)]=\int^{t_{f}}_t H\,dt=\kappa\int^{\phi_{f}}_\phi(H/\dot{\phi})d\phi,
\end{equation}
where $N$ is the value of the number of $e$-folds at the cosmological time ``t" and  $N_f$ denotes the number of  $e$-folds at the end of inflation. In addition, in order to analyze  the inflationary epoch, it is also useful to define the following
slow-roll  parameters \cite{K1,K2}
\begin{equation}
    \epsilon=-\frac{\dot{H}}{H^2},\,\,\,\,\,\,
    \eta=\frac{\dot{\epsilon}}{H\epsilon},\,\,\,\,\,\,\,\,\,
    s=\frac{\dot{c}_s}{Hc_s}\label{s},
\end{equation}
in which the parameters  $\epsilon$, $\eta$ and $s$ are much less the unity. Thus, the inflationary scenario occurs when  the $\epsilon$ parameter is less than unity, which is equivalent to the fact that the  $\ddot{a}>0$ (accelerated phase). Besides,
the inflationary epoch ends when the parameter $\epsilon=1$ or equivalently $\ddot{a}=0$,and also when the $\eta$ parameter approach unity. However, if the speed of sound does not change significantly  with respect to the time during inflation, the parameter $s$ could be  much smaller than unity. 

In this context, the  equations of motion assuming the set of slow-roll conditions are reduced to \cite{K1,K2}
\begin{equation}
\label{KleinGordon}
    H^2\simeq \frac{\kappa}{3}\,V,\,\,\,\,\,\,\,\,\,\,\mbox{and}\,\,\,\,\,\,\,\,\,\,\,\,\,\,3H\dot{\phi}\,(g\dot{\phi}^2+1)+V_\phi\simeq0,
\end{equation}
respectively. Here the equation for the scalar field in the slow roll approximation  differs from the canonical case due to the existence of the term $3Hg\dot{\phi}^3$. In this way, we can observe two limits; in the case in which  $1\gg g\dot{\phi}^2$ and it corresponds to the standard canonical theory and the situation in which $ g\dot{\phi}^2\gg 1$ predominates and then the dynamics of the canonical theory is modified. 

On the other hand, for a non canonical theory  the power spectrum of the  curvature perturbations $A_s$, together with the spectral index $n_s$ were obtained in Refs. \cite{K1,K2,DeFelice:2013ar}.
In this context, following Ref.\cite{DeFelice:2013ar} we can write that the second-order action for the metric perturbation $\mathcal{R}$ becomes
\begin{equation}
\mathcal{S}_2=\int\,dt\,d^3x\,a^3\,Q\left[\dot{\mathcal{R}}^2-\frac{c_s^2}{a^2}(\partial \mathcal{R})^2\right],
\label{ac2}
\end{equation}
where the quantity $Q$ is defined as $Q=X(1+6gX)/H^2$. Thus from the action given by Eq.(\ref{ac2}) we have \cite{DeFelice:2013ar}
\begin{equation}
\label{As}
    A_s=\frac{\kappa}{8\pi^2}\frac{H^2}{c_s\epsilon},
\end{equation}
and
\begin{equation}
\label{nsm1}
    n_s-1=\frac{d A_s}{d\ln k}\simeq-(2\epsilon+\eta+s),
\end{equation}
respectively.

Besides, the generation of tensor perturbation $A_t$ during the inflationary period is not modified in the framework of non canonical theory  and then its expression coincides with the canonical theory,  with which the tensor perturbation  $A_t=(2\kappa H^2/\pi^2)$.  However, the quantity associated to the  tensor to scalar ratio $r$ under the slow roll approximation  is modified by the factor $c_s$ and it becomes\cite{K1,K2,DeFelice:2013ar}
\begin{equation}
\label{r}
    r=\frac{A_t}{A_s}= 16c_s\epsilon.
\end{equation}
Thus, in the frame of a canonical theory in which the speed of sound $c_s=1$, the tensor to scalar ratio is reduced to the standard expression $r=16\epsilon$. In relation to the scalar spectral index obtained in Ref.\cite{Hwang:2005hb} which is defined as $n_s=4\epsilon_1-2\epsilon_2-2\epsilon_4$, where the slow roll parameters are given by $\epsilon_1=\dot{H}/H^2=-\epsilon$, $\epsilon_2=\ddot{\phi}/(H\dot{\phi})$ and $\epsilon_4=\dot{E}/(2HE)$ with $E=1+2gX$, only coincides with the spectral index given by Eq.(\ref{nsm1}) if the slow roll parameter $s=0$. This difference in the scalar spectral indexes   occurs due to  the definition of the slow roll parameter $\eta$ defined by Eq.(\ref{s}), since this parameter  can be rewritten as $\eta=-2\epsilon_1+2\epsilon_2+2\epsilon_4$ using the slow roll equations given by Eq.(\ref{KleinGordon}). Thus considering $s=0$ in Eq.(\ref{nsm1}), we find that  both scalar spectral indexes are equivalent.  

In the following we will study the reconstruction of inflation in the framework of  a non canonical theory. In this sense, for the reconstruction  the background variables we will determine  the effective potential $V$ together with the coupling function $g$ in terms of the scalar field $\phi$ i.e., $V(\phi)$ and $g(\phi)$, respectively.

\section{General reconstruction}\label{reconstruction}

In this section, we will  realize  the procedure to the reconstruction of the background variables, such as; the effective potential and the coupling function $g$  as a function of the scalar field, in the context of a non canonical theory, considering the scalar spectral index $n_s$ and the speed of sound $c_s$ in terms of the number of $e-$ folds $N$.   In this context, we will rewrite the spectral index $n_s$ and the coupling function $g$
as a function of the potential and of the speed of sound in terms of the number of $e-$ folds $N$ and its derivatives with respect to  $N$, i.e., $V_N=\partial V/\partial N$, $V_{NN}=\partial^2 V/\partial N^2$, ${c_{s}}_{N}=\partial c_s/\partial N$, etc.
In this way, from these  quantities and giving 
the spectral index $n_s=n_s(N)$ and the speed of sound $c_s=c_s(N)$, we should find the effective potential $V=V(N)$ together with the coupling function $g=g(N)$ as a function of the number of $e-$ folds $N$. Subsequently, from the relation between the number $N$ and the scalar field $\phi$ given by the differential equation (\ref{deltaN}), we should obtain the number $N$ in terms of $\phi$, i.e., the relation $N=N(\phi)$ and then we should rebuild the effective potential $V(\phi)$ and the function $g(\phi)$ associated to the non-linear kinetic term $gX^2$ of the Lagrangian density given by Eq.(\ref{p}).

In this sense, we begin by rewriting  the function coupling and the scalar spectral index  given by Eqs. (\ref{gf}) and (\ref{nsm1}) in terms of the number $N$.
In this form, the slow roll parameters $\epsilon$ and $\eta$ can be rewritten as a function of the number of $e-$ folds $N$ using the fact that 
\begin{equation}
V_\phi=\frac{\partial V}{\partial \phi}=\frac{\partial V}{\partial N}\,\frac{\partial N}{\partial \phi}=V_N\,N_\phi=-V_N\left(\frac{H}{\dot{\phi}}\right),\,\,\,\,\,\mbox{where}\,\,\,\,\,
N_\phi=-\frac{H}{\dot\phi}.\label{vf5}
\end{equation}
Here we have considered that the quantity $dN=-Hdt=-Hd\phi/\dot{\phi}$ \cite{Chiba,Herrera:2018cgi}.

Thus, we find that the first slow roll parameter $\epsilon$ can be written as
\begin{equation}
    \epsilon=-\frac{1}{2}\left(\frac{V_\phi}{V}\right)\left(\frac{\dot{\phi}}{H}\right)=\frac{1}{2}\left(\frac{V_N}{V}\right).
\label{e1}
\end{equation}

Similarly, the  slow roll parameter $\eta$ defined as $\eta=\frac{\dot{\epsilon}}{H\epsilon}=-\frac{d\ln \epsilon}{dN}$ becomes 
\begin{equation}
   \eta=-\frac{V_{NN}}{V_N}+\frac{V_N}{V}, 
\end{equation}
and the slow roll parameter associated to speed of sound $s$ results
\begin{equation}
s=\frac{\dot{c}_s}{Hc_s}=-\frac{d \ln c_s}{dN}.
\end{equation}

In this form, using the above  equations, we can rewrite the scalar spectral index given by Eq.(\ref{nsm1}) in terms of the effective potential and the speed of sound together with  its derivatives with respect to $N$ as
\begin{equation}
\label{EqnsVN}
    n_s-1\simeq -2\frac{V_N}{V}+\frac{V_{NN}}{V_N}+\frac{d \ln c_s}{dN}.
\end{equation}
Thus, we obtain that the effective potential in  terms of the number of $e-$ folds can be written as
\begin{equation}
V(N)=V=\left[-\int \left(\exp\int\left[(n_s-1)-\frac{d\ln c_s}{dN}\right]dN\right)dN\right]^{-1}.
\label{VV5a}
\end{equation}
In the same form,  for the coupling function $g$ we have
\begin{equation}
 \label{er1}
   g(N)= g=\left(\frac{1-c_s^2}{3c_s^2-1}\right)\frac{1}{\dot\phi^2}=3\left(\frac{1-c_s^2}{3c_s^2-1}\right)
\left[1+\left(\frac{1-c_s^2}{3c_s^2-1}\right)\right]\,\frac{1}{V_N},
\end{equation}
or equivalently
$$
g(N)=3\left(\frac{1-c_s^2}{3c_s^2-1}\right)
\left[1+\left(\frac{1-c_s^2}{3c_s^2-1}\right)\right]\,\frac{\exp\left(\int\left[(1-n_s)+\frac{d\ln c_s}{dN}\right]dN\right)}{V^2}\,\,=
$$
\begin{equation}
\label{gVn}
3\left(\frac{1-c_s^2}{3c_s^2-1}\right)
\left[1+\left(\frac{1-c_s^2}{3c_s^2-1}\right)\right]\,\exp\left(\int\left[(1-n_s)+\frac{d\ln c_s}{dN}\right]dN\right)\left[\int \left(\exp\int\left[(n_s-1)-\frac{d\ln c_s}{dN}\right]dN\right)dN\right]^{2} ,
\end{equation}
where we have used Eqs.(\ref{KleinGordon}) and (\ref{vf5}), respectively.
Here we note that in the special case in which the speed of sound is a constant equal to unity, Eqs.(\ref{VV5a}) and (\ref{gVn}) reduce to the standard reconstruction in the framework of a canonical theory, i.e., $V(N)$ corresponds to Ref.(\cite{Chiba}) and $g=0$.

In this way, we mention that the Eqs.(\ref{VV5a}) and (\ref{gVn}) are the essential equations to obtain the effective potential and the coupling function in terms of the number of $e-$  folds $N$. In this sense, in order to obtain these background variables as a function of $N$, we need to give the parametrization of the scalar spectrum index $n_s=n_s(N)$ together with the parametrization of the speed of sound $c_s=c_s(N)$.

For the power spectrum of the curvature perturbations from Eq.(\ref{As}) we have
\begin{equation}
A_s\simeq\frac{\kappa^2}{12\pi^2}\,\left[\frac{V^2}{c_s\,V_N}\right],
\end{equation}
and for the tensor to scalar ratio in terms of the potential and speed of sound results
\begin{equation}
r=8c_s\,\left(\frac{V}{V_N}\right).
\label{rr2}
\end{equation}

Additionally, in order to find the relation between the number of $e-$ folds $N$ and the scalar field $\phi$ we get
\begin{equation}
\,\left[\frac{V_N\,(3c_s^2-1)} {V\,c_s^2}\right]^{1/2}dN=\sqrt{2\kappa}\,d\phi,
\label{Rr2}
\end{equation}
where  we have used  Eqs.(\ref{gf}) and (\ref{deltaN}), respectively.

Here, we mention that the system of equations given by Eqs.(\ref{VV5a}), (\ref{gVn}) and (\ref{Rr2}) are the basic equations to rebuild the effective potential $V$ and the coupling function $g$, in terms of the scalar field $\phi$, i.e., $V=V(\phi)$ and $g=g(\phi)$, respectively.

In the following, we will assume a specific example for the scalar spectral index as well as for the speed of sound in terms of the number of $e-$ folds $N$, to reconstruct the effective potential $V(\phi)$ and the coupling function $g(\phi)$, respectively.

\section{Reconstruction from $n_s=n_s(N)$ and $c_s=c_s(N)$}\label{Rec2}

In this section we will apply the methodology previously described, assuming a parametrization for both the scalar spectral index and the sound of speed in terms of the number of $e-$ folds $N$, in order to rebuild the background variables; the effective potential $V$ and the coupling function $g$ as a function of the scalar field $\phi$. 

In this sense, we can assume the simplest parametrization or attractor for the scalar spectral index $n_s$ as a function of $N$ given by \cite{Chiba:2015zpa}
\begin{equation}
n_s(N)=1-\frac{2}{N},\,\,\,\,\,\,\mbox{in which }\,\,\,\,\,\,\,\,N\neq 0,
\label{nn}
\end{equation}
and for  
the speed of sound propagation of the scalar perturbations we consider the parametrization  
\begin{equation}
\label{csN}
c_s(N)=c_{s_{f}}\left(\frac{N_f}{N}\right)^{\beta}=\tilde{c}_{s_f}N^{-\beta},
\end{equation}
where the quantity  $c_{s_f}$ is  a constant and it corresponds to the value of the speed of sound propagation at the end of inflation, the quantity  $\tilde{c}_{s_f}$ is defined as $\tilde{c}_{s_f}=c_{s_f}N_f^{\beta}$ and the power $\beta$ is a constant. In relation to the parametrization of the speed of sound given by Eq.(\ref{csN}) in terms of the number of $e-$folds $N$, we mention that a study associated to the reconstruction of an inflationary stage using this  parametrization  does not exist in the literature. In this form, the motivation to consider the simplest parametrization of the speed of sound $c_s=c_s(N)$ described by Eq.(\ref{csN}) comes from 
 the speed of sound analyzed in Ref.\cite{Shi:2021tmn} in which a power law type dependence on  time 
$c_s\propto (-\tau)^s$ is considered  or with the energy density $c_s^2\propto \rho^{1-2n}$ as in \cite{Magueijo:2008pm}.

In this context, the present work is the first step towards that direction using this physical parameter and this dependence with the number of $e-$folds during inflation, in order to rebuild the background variables during the early universe.

By assuming Eq.(\ref{csN}) and  as the speed propagation satisfies the relation $1\geqslant c_s^2 >1/3$ (see Eq.(\ref{gf})), then we find that the upper and lower bounds for $\beta$ becomes
\begin{equation}
\label{beta1}
    \frac{\ln (c_{s_f})}{\ln( N/N_f)}\leq\beta<\frac{\ln (3^{1/2}\,c_{s_f})}{\ln (N/N_f)}.
\end{equation}
However, we observe that as  the number of $e-$folds $N\sim\mathcal{O}(10^2)\gg N_f\sim\mathcal{O}(1)$  and the speed of sound  $c_{s_f}\sim\mathcal{O}(1)$ (since $1\geqslant c_{s_f}^2 >1/3$), then from Eq.(\ref{beta1}) we have that the parameter $\beta$ will always be less 
 than unity i.e., $\beta<1$.  

On the other hand, in order to find the potential in terms of the number of $e-$ folds $N$, we replace Eqs.(\ref{nn}) and (\ref{csN}) into Eq.(\ref{VV5a}) resulting 
\begin{equation}
\label{V(N)}
    V(N)=\left(C+\frac{\alpha}{1-\beta}N^{\beta-1}\right)^{-1},
\end{equation}
where $\alpha$ and $C$  are integration  constants and  these constants have units of $M_p^{-4}$.  Similarly, from Eq.(\ref{gVn}) we find that the coupling parameter $g$ as a function of the number of $e$-folds becomes
\begin{equation}
\label{g(N)}
    g(N)=\frac{3}{\alpha}\left(\frac{1-\tilde{c}_{s_f}^{\,2} N^{-2\beta}}{1-3\,\tilde{c}_{s_f}^{\,2} N^{-2\beta}} \right)^2\left(C+\frac{\alpha}{1-\beta}N^{\beta-1} \right)^2N^{2-\beta}.
\end{equation}
Therefore, to obtain the reconstruction of the effective potential $V(\phi)$ and the coupling function $g(\phi)$, we need to find the relation between the number of $e-$ folds $N$  and the scalar field $\phi$  given by Eq.(\ref{Rr2}).

In order to find the relation between the number of $e-$ folds $N$ and the scalar field $\phi$ i.e., $N=N(\phi)$ from Eq.(\ref{Rr2}),   we need to  replace the  potential $V(N)$ and $c_s(N)$ given by Eqs.(\ref{csN}) and (\ref{V(N)}) into (\ref{Rr2}). However, we cannot analytically solve  this equation. Thus,     
 to solve the Eq.(\ref{Rr2}), we can assume that  during the inflationary epoch  the integration constant $C$ is a positive quantity and   $C\gg \alpha N^{\beta-1}/[(1-\beta)]$. Under this approximation,  we find that the effective potential as a function of the number $N$ given by Eq.(\ref{V(N)}) can be  approximate to 
\begin{equation}
\label{V(N)aprox}
V(N)\simeq C^{-1}\left(1-\frac{\alpha}{(1-\beta)C} N^{\beta-1}+\dotsb \right).
\end{equation}
In relation to the approximation in which $C\gg \alpha N^{\beta-1}/[(1-\beta)]$, we find that  the  potential given by Eq.(\ref{V(N)}) is reduced to the potential  Eq.(\ref{V(N)aprox}) and then  considering  the slow roll regime where $H^2\propto V$, we have that the early universe undergoes a nearly exponential expansion supported by this condition. In this sense, this type of expansion of the universe has been widely studied in the literature, see Refs.\cite{1A, Barrow:1990vx}. 

In this form, the  differential equation associated to the number of $e-$folds and the scalar field given by Eq.(\ref{Rr2}) is reduced to  
\begin{equation}
    \left(\frac{2\kappa C}{\alpha}\right)^{1/2} \simeq\left(\frac{3\tilde{c}_{s_f}^{\,2}N^{-2\beta}-1}{\tilde{c}_{s_f}^{\,2}N^{-2\beta}}\right)^{1/2}\left(1+\frac{\alpha}{2(1-\beta)C}N^{\beta-1}\right)N^{\beta/2-1}N_\phi.
\end{equation}
Solving this differential equation we obtain that the number of $e$-folds as a function of the field $\phi$ can be written as
\begin{equation}
\label{N}
N(\phi)=N=\mathcal{F}^{-1}(\phi),
\end{equation}
where the function $\mathcal{F}^{-1}(\phi)$ represents the inverse function of 
\begin{equation}
{\mathcal{F}}(\phi)=2\sqrt{3}\left(\sqrt{\frac{8\kappa C^3}{\alpha}}(1-\beta)\phi+\gamma\right)^{\beta/2}\left[\mathcal{F}_1(\phi)+\mathcal{F}_2(\phi)\right],
\end{equation}
where the functions $\mathcal{F}_1(\phi)$ and $\mathcal{F}_2(\phi)$ are defined as
\begin{eqnarray}
    \mathcal{F}_1(\phi)&=&\frac{2(1-\beta)C}{\beta}\,\,_2F_1\left[-\frac{1}{2},\frac{1}{4},\frac{5}{4},\frac{\left(\sqrt{\frac{8\kappa C^3}{\alpha}}(1-\beta)\phi+\gamma\right)^{2\beta}}{3\tilde{c}_{s_f}^{\,2}}\right],\,\,\,\,\,\,\,\,\,\,\,\mbox{and}\\
    \mathcal{F}_2(\phi)&=&\frac{\alpha \left(\sqrt{\frac{8\kappa C^3}{\alpha}}(1-\beta)\phi+\gamma\right)^{\beta-1}}{3\beta-2}\,\,_2F_1\left[-\frac{1}{2},\frac{1}{4}\left(3-\frac{2}{\beta}\right),\frac{1}{4}\left(7-\frac{2}{\beta}\right),\frac{\left(\sqrt{\frac{8\kappa C^3}{\alpha}}(1-\beta)\phi+\gamma\right)^{2\beta}}{3\tilde{c}_{s_f}^{\,2}}\right],
\end{eqnarray}
respectively. Besides, the parameter $\gamma$ is an  integration constant  and the quantity $\,_2F_1$ corresponds to  the hypergeometric function\cite{Libro}.

In this form, replacing the number of $e-$ folds in terms of the scalar field given by Eq.(\ref{N}) into the  potential (\ref{V(N)aprox}), we find that the reconstruction of the effective  potential as a function of the scalar field becomes
\begin{equation}
    V(\phi)\simeq 
C^{-1}\left(1-\frac{\alpha}{(1-\beta)C} \left[\mathcal{F}^{-1}(\phi)\right]^{\beta-1}\right).\label{V(phi)}
\end{equation}
Analogously, for the coupling function $g(\phi)$, we obtain that the reconstruction for this background variable results
\begin{equation}
g(\phi)\simeq\frac{3C^2}{\alpha}\left(\frac{1-\tilde{c}_{s_f}^{\,2}\left[\mathcal{F}^{-1}(\phi)\right]^{-2\beta}}{3\,\tilde{c}_{s_f}^{\,2}\left[\mathcal{F}^{-1}(\phi)\right]^{-2\beta}-1}\right)\left[1+\left(\frac{1-\tilde{c}_{s_f}^{\,2}\left[\mathcal{F}^{-1}(\phi)\right]^{-2\beta}}{3\,\tilde{c}_{s_f}^{\,2}\left[\mathcal{F}^{-1}(\phi)\right]^{-2\beta}-1}\right)\right]\left[\mathcal{F}^{-1}(\phi)\right]^{2-\beta}\label{g(phi)}.
\end{equation}

Additionally, we have that the speed of sound as a function of the scalar field can be written as
\begin{equation}
c_s(\phi)\simeq\tilde{c}_{s_f}\left[\mathcal{F}^{-1}(\phi)\right]^{-\beta},\label{cs(phi)}
\end{equation}
here in the reconstruction of $c_s(\phi)$, we have considered Eq.(\ref{csN}).

\begin{figure}[h]
  \centering
  \begin{subfigure}{0.4\textwidth}
    \centering
    \includegraphics[width=\linewidth]{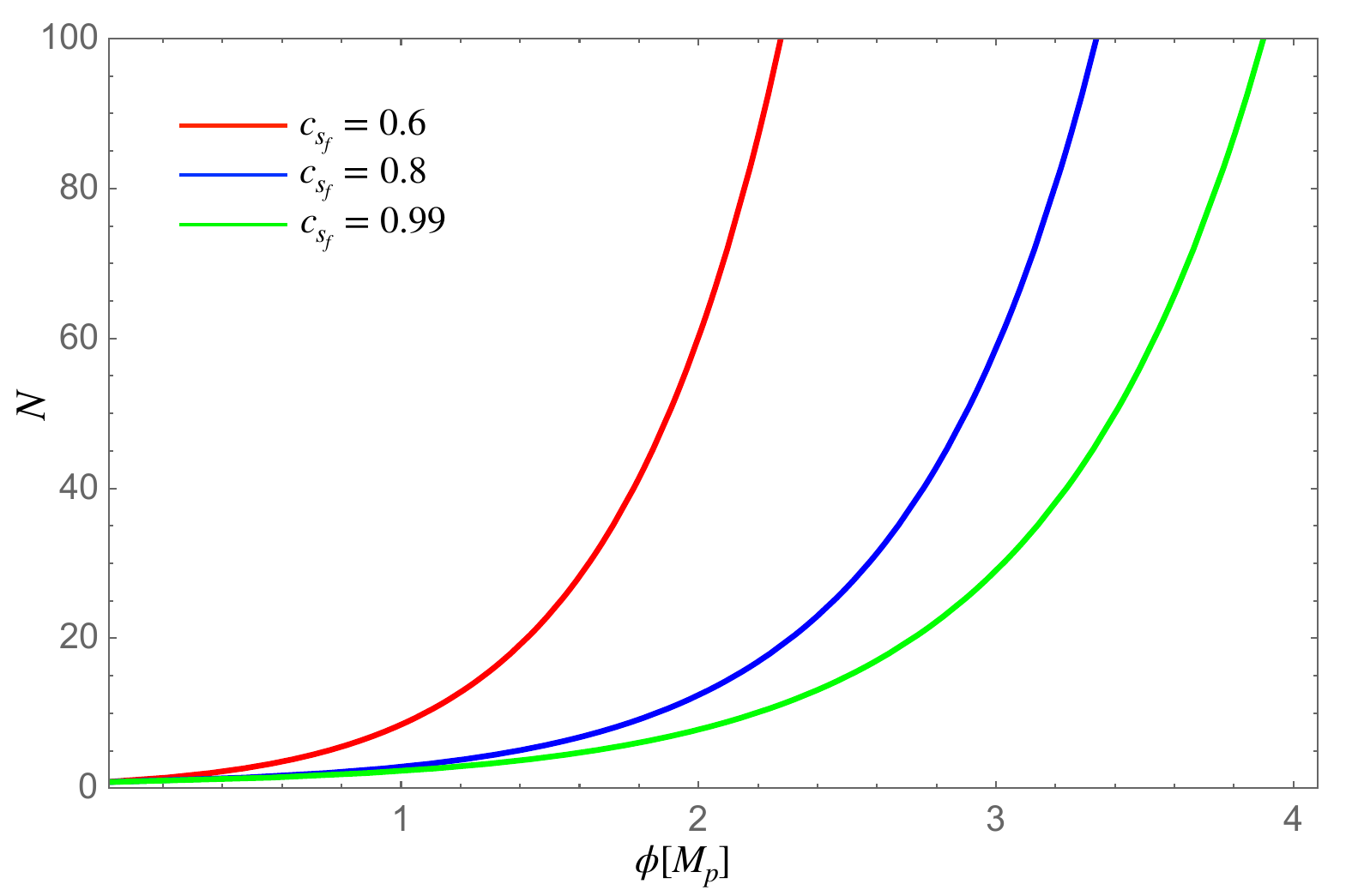}
    \end{subfigure} 
  
  \begin{subfigure}{0.42\textwidth}
    \centering
    \includegraphics[width=\linewidth]{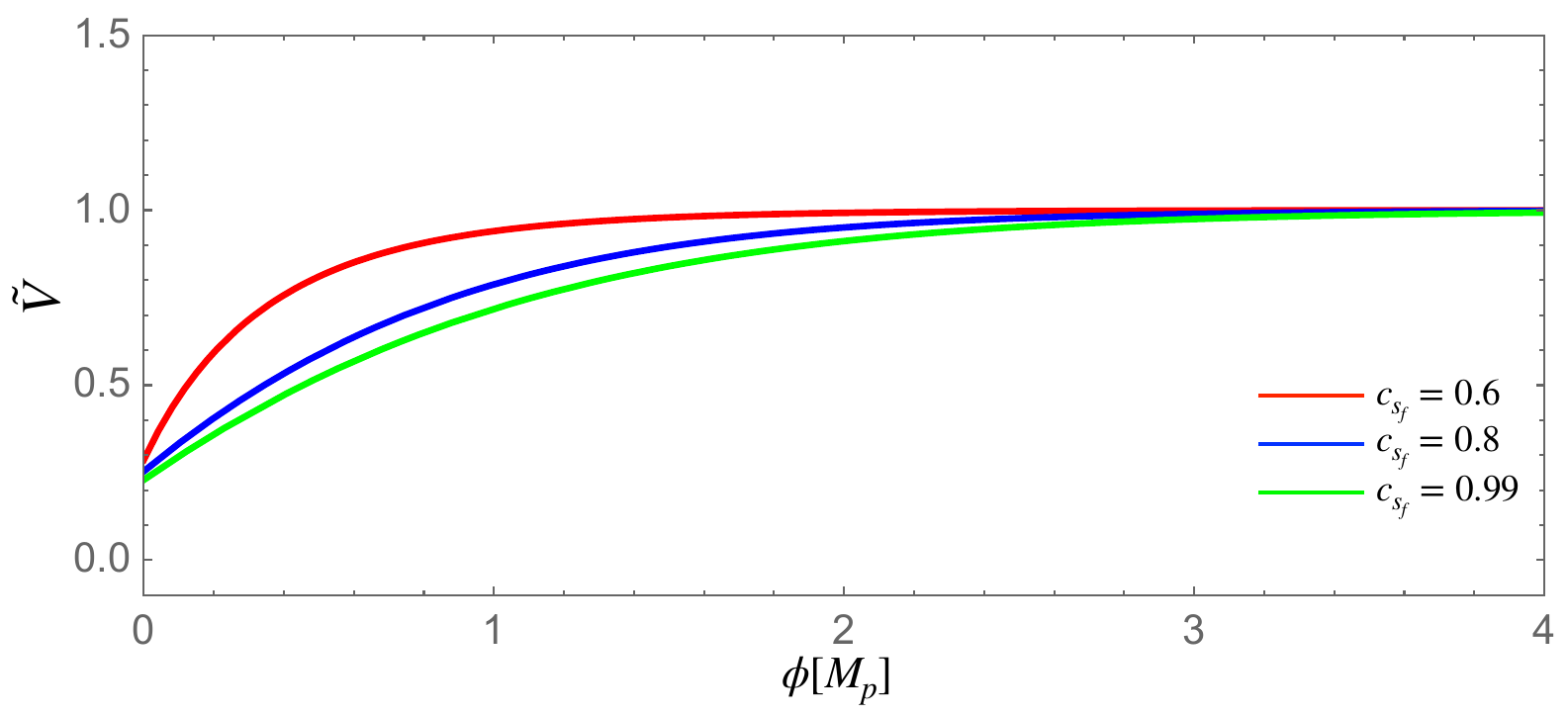}
    \end{subfigure}%
  \caption{ The behavior of the number of $e-$ folds $N$ on the scalar field $\phi$ (upper panel) and the reconstructed effective potential dimensionless $\tilde{V}=C\,V$ versus the scalar field $\phi$ (lower panel), for different values of the speed of sound $c_{s_{f}}$ at the end of inflationary epoch. In these plots we have used the values $C\simeq 3.04\times 10^{10}\kappa^{2}$, $\alpha\simeq 1.80\times 10^{10}\kappa^{2}$ and $N_f=0.8$.  }
\label{Fig1}
\end{figure}
\begin{figure}[h]
  \begin{subfigure}{0.6\textwidth}
    \includegraphics[width=0.5\linewidth]{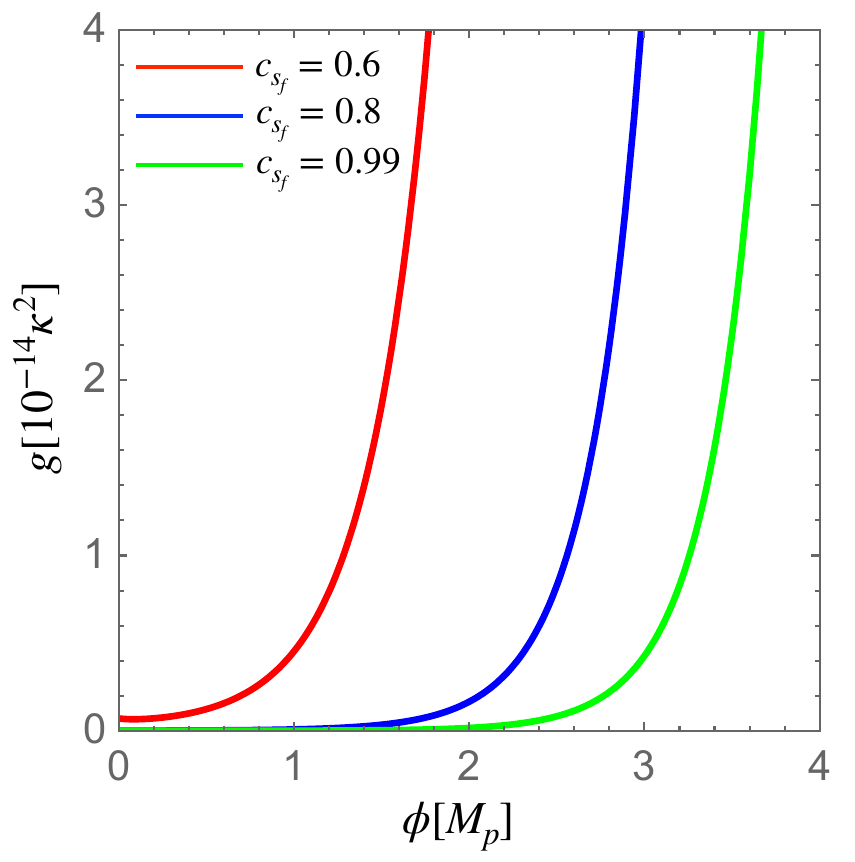}
    \end{subfigure} \hfill
   \begin{subfigure}{0.7\textwidth}
    \includegraphics[width=0.6\linewidth]{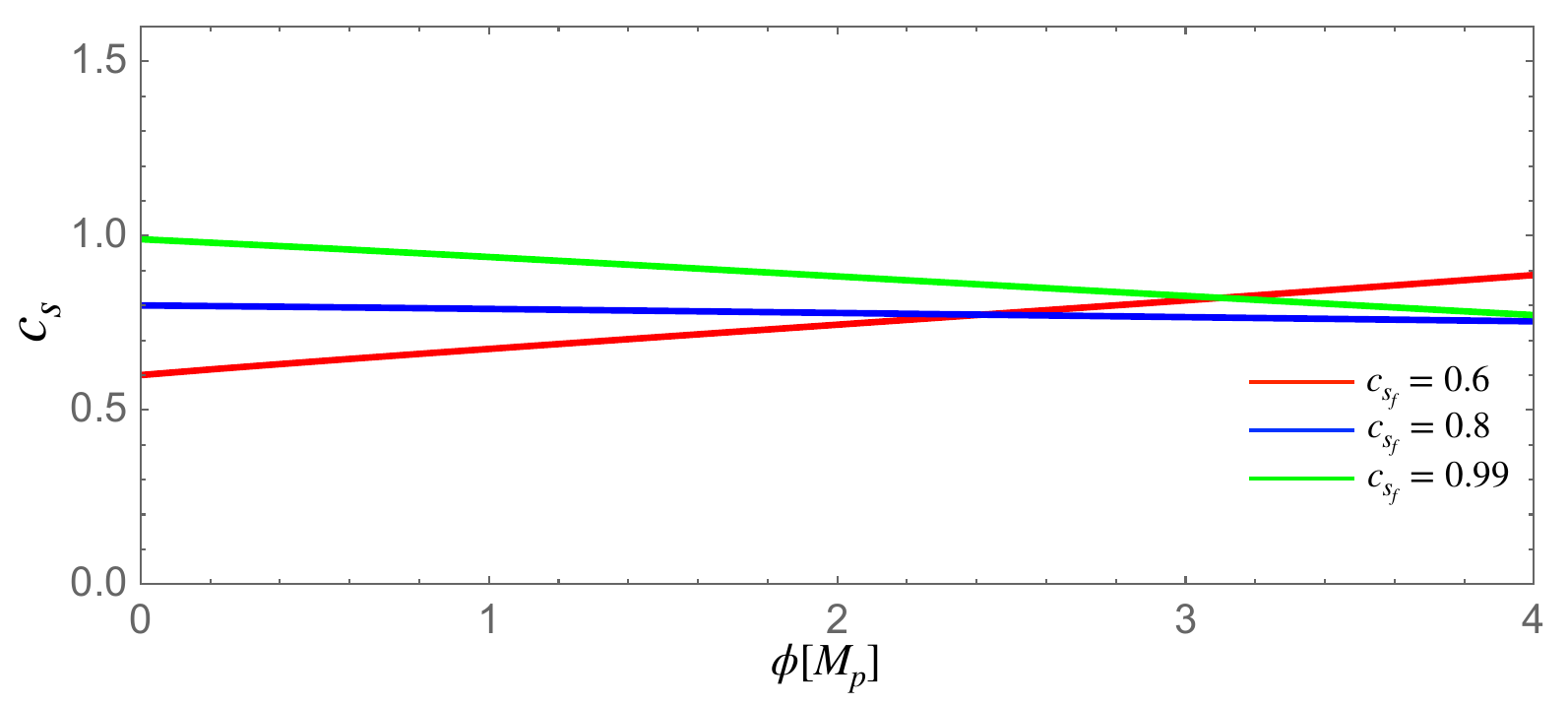}
    \end{subfigure} 
  \caption{ The upper panel shows the reconstruction of the coupling function $g$ as a function of the scalar field  and the lower panel shows the speed of sound in terms of the scalar field i.e., $c_{s}(\phi)$. As before we have considered different values of the speed of sound at the end of inflation $c_{s_{f}}$. Besides, we have used the values $C\simeq 3.04\times 10^{10}\kappa^{2}$, $\alpha\simeq 1.80\times 10^{10}\kappa^{2}$ and $N_f=0.8$, respectively.}
  \label{Fig2}
\end{figure}

\begin{figure}[h]
    \centering
    \includegraphics[width=1\linewidth]{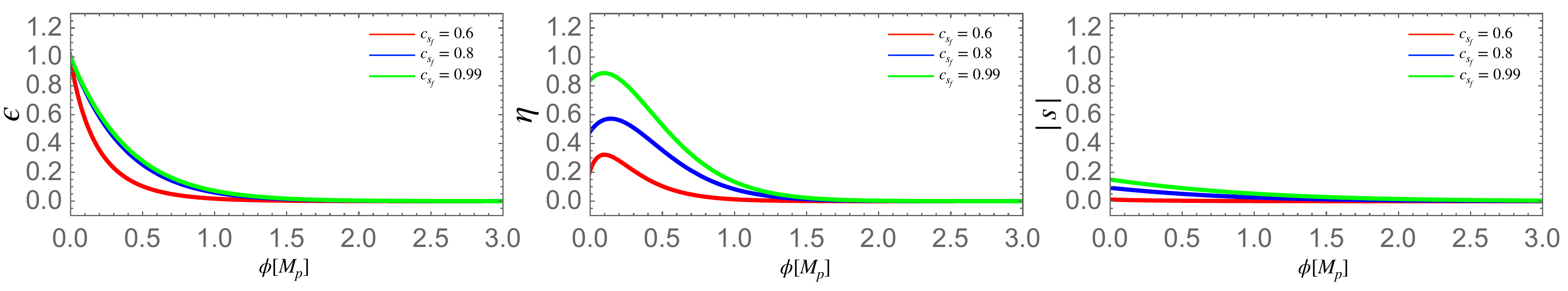}
    \caption{The graphs show the evolution of 
slow roll parameters as a function of the scalar field $\phi$, for some values of the speed of sound $c_{s_{f}}$ at the end of inflation
epoch. As before we have considered different values of the speed of sound at the end of inflation $c_{s_{f}}$. Besides, we have used the values $C\simeq 3.04\times 10^{10}\kappa^{2}$, $\alpha\simeq 1.80\times 10^{10}\kappa^{2}$ and $N_f=0.8$, respectively.} 
    \label{fig_SlowRoll}
\end{figure}

In Fig.\ref{Fig1}, the upper panel shows the evolution of the number of $e-$ folds $N$ versus the scalar field $\phi$, while the lower panel shows the reconstructed effective potential $\tilde{V}$ in terms of the scalar field. In both panels, we consider the situation in which the speed sound at the end of inflation $c_{s_{f}}$ has three different values.  Besides, in these plots we have assumed the values of  $C\simeq 3.04\times 10^{10}\kappa^{2}$, $\alpha\simeq 1.80\times 10^{10}\kappa^{2}$ and $N_f=0.8$, respectively. In order to write down values of the number of $e-$ folds $N(\phi)$ and the effective potential $\tilde{V}(\phi)$, we have used Eqs.(\ref{N}) and (\ref{V(phi)}), respectively. From the upper panel we observe that the end of the inflationary epoch in which $N_{f}\sim\mathcal{O}(1)$ the   scalar field takes the values  $\phi_f\sim\mathcal{O}(0)$. Also,  we observe that the number of $e-$folds $N$ takes the values $N\sim 60$  when the scalar field is approximately  $\phi\sim\mathcal{O}(M_p)$.
From the lower panel we note that for values of $\phi>3M_p$, the reconstructed potential becomes constant equal to $V\sim C$ and independent of the value of the sound seed at the end of inflation $c_{s_{f}}$.

In Fig.\ref{Fig2}, the upper panel shows the reconstruction of the coupling function versus the scalar field $\phi$, while the lower panel shows the reconstructed speed of sound  in terms of the scalar field.
As before, in both panels we consider the case in which the speed sound at the end of inflation $c_{s_{f}}$ has three different values.  Also, in these plots we have assumed the values of  $C\simeq 3.04\times 10^{10}\kappa^{2}$, $\alpha\simeq 1.80\times 10^{10}\kappa^{2}$ and $N_f=0.8$, respectively. From Fig.\ref{Fig2}  we note that the coupling function increases very fast for large $N$ (see relation between $\phi$ and $N$ of Fig.\ref{Fig1}). Besides, from the lower panel we observe that the speed of sound presents a small slope in relation to the scalar field. Here we note that during the inflationary scenario the cases in which $c_{s_{f}}=0.99$ (green line) or $c_{s_{f}}=0.6$ (blue line) the evolution of the speed $c_s(\phi)$ increases up to the specified value of $c_{s_{f}}$. However, for the situation in which $c_{s_{f}}=0.99$ (red line) this dependence with the scalar field decreases when the field moves from $N\sim\mathcal{O}(10^2)$ to the end of inflation where $N_{f}\sim\mathcal{O}(1)$.

Figure \ref{fig_SlowRoll} portrays the evolution of the different slow roll parameters in terms of the scalar field defined by Eq.(\ref{s}). Here we note that these parameters are less than unity, in order to satisfy the slow- roll approximation during the inflationary scenario. Besides, we observe from Fig.(\ref{Fig2}) (lower panel) that  the sound of speed $c_s=c_s(\phi)$ exhibits a small variation with respect to  the scalar field, then this suggests that the parameter $s$  tends toward zero, as shown in the right panel of 
Fig.\ref{fig_SlowRoll}.

On the other hand,  to constrain  the integration constant $\alpha$, we can consider  the expression of the scalar power spectrum given by Eq.(\ref{As}). Thus,  using the Eqs.(\ref{As})  and (\ref{csN}) we obtain that the power spectrum $A_s$ is given by

\begin{equation}
    A_s\simeq\frac{\kappa^2}{12\alpha \pi^2 \tilde{c}_{s_f}}N^2\left[1-\frac{2\alpha N^{\beta-1}}{(1-\beta)C}\right],
\label{AS}
\end{equation}
and then the integration constant $\alpha$ becomes
\begin{equation}
    \alpha=\frac{\kappa^2\,N^2}{12 \pi^2 \tilde{c}_{s_f}\,A_s}
\left[1+\frac{\kappa^2N^2}{6\pi^2A_sC\,c_{s_f}N_f\,x}\right]^{-1},\,\,\,\,\,\,\mbox{where}\,
\,\,\,\,\,\,\,x=(1-\beta)\left(\frac{N}{N_f}\right)^{(1-\beta)}.
\label{x1}
\end{equation}

Besides, from the effective potential given by Eq.(\ref{V(N)aprox}) together with Eq.(\ref{rr2}), we obtain that  the tensor to scalar ratio as a function of the number of $e$-folds $N$ yields
\begin{equation}
\label{r(n)}
    r\simeq \frac{8 \alpha\,\tilde{c}_{s_f}}{C\,N^2}\left[1-\frac{\alpha}{(1-\beta)C}N^{\beta-1}\right]^{-1}.
\end{equation}
In this way, from the tensor to scalar ratio (\ref{r(n)}) and considering Eq.(\ref{x1}),
 we find  an equation for the quantity $x$ (defined in Eq.(\ref{x1})) given by
\begin{equation}
x=(1-\beta)\left(\frac{N}{N_f}\right)^{(1-\beta)}=\frac{\alpha_0\alpha_1}{(1-\alpha_1C)}=\alpha_2\,\,\,\,\,\mbox{with}\,\,\,\,\,\alpha_1 C\neq 1.\label{bb}
\end{equation}
 Here, 
 the quantities $\alpha_0$ and $\alpha_1$ become
\begin{equation}
\alpha_0=\frac{\tilde{\alpha}_0}{N_f}=\frac{\kappa^2\,N^2}{12\pi^2\,A_s\,c_{s_f}N_f},\,\,\,\,\,\,\mbox{and}\,\,\,\,\,\,\alpha_1=\frac{3\pi^2\,A_s\,r}{2\kappa^2}\,, 
\end{equation}
respectively. 
In this form, the solution of Eq.(\ref{bb})  for the parameter $\beta$ can be written in terms of  the observational parameters $A_s$, $r$ and the number of $e-$ folds $N$  as 
\begin{equation}
\label{exactbeta}
\beta=1-\frac{\mbox{ProductLog}\left[\alpha_2 \,\ln(N/N_f)\,\right]}{\ln(N/N_f)}.
\end{equation}
Here the ProductLog function corresponds to a product logarithm, also
called the Omega function or Lambert W function,  see e.g., Refs.\cite{87,88}.

Now inserting Eq.(\ref{exactbeta}) into Eq.(\ref{beta1}), we find that the second integration constant $C$ is constrained by an upper and lower bounds from the observational parameters $A_s$, $r$ and the number of $e-$ folds $N$ together with the parameters at the end of inflation $c_{s_{f}}$ and  $N_f$ as
\begin{equation}
\label{Cinequality}
    \frac{F_1-1}{\alpha_1\,F_1}< C \leq \frac{F_2-1}{\alpha_1\,F_2}  ,
\end{equation}
where the quantities $F_1$ and $F_2$ are defined as 
\begin{equation}
    F_1=\frac{(N/N_f)}{\sqrt{3}\alpha_0\alpha_1 c_{s_f}}\left[1-\frac{\ln (\sqrt{3}c_{s_f})}{\ln (N/N_f)}\right]\,\,\,\,\,\,\mbox{and}\,\,\,\,\,\,F_2=\frac{(N/N_f)}{\alpha_0\alpha_1 c_{s_f}}\left[1-\frac{\ln (c_{s_f})}{\ln (N/N_f)}\right].
\end{equation}

On the other hand, in order to obtain the number of $e-$folds at the end of inflation $N_f$, we can consider  the slow roll parameter $\epsilon$ defined by  Eq.(\ref{e1}) together with the reconstructed potential (\ref{V(N)aprox}). In this way,  assuming  that inflation ends when $\epsilon(N=N_f)=1$ (or equivalently $\ddot{a}=0$), we have that
\begin{equation}
\label{EqNf}
    N_f^3-p_0\,N_f-\frac{2p_0^2}{(1-\beta)}=0,\,\,\,\,\,\,\mbox{where}\,\,\,\,\,\,\,\,\,p_0=\frac{\tilde{\alpha}_0}{2(2/\alpha_1-C)}.
\end{equation}
Here the parameter $p_0$ is a function of the $N_f$,  and then the Eq.(\ref{EqNf}) does not have an analytical solution. However, we can solve the Eq.(\ref{EqNf})  numerically.  Thus, 
in particular considering the values of  
$N=60$, $A_s= 2.2\times10^{-9}$, $r=0.04$, $c_{s_f}=1$ and assuming the upper bound of Eq.(\ref{Cinequality}) for $C$ (that depends of $N_f$),  then  numerically we find that the number of $e-$folds at the end of inflation from Eq.(\ref{EqNf}) becomes $N_f\simeq 5.08$   and for the case in which  $c_{s_f}=0.6$  we have $N_f\simeq 6.96$. Now for the case in which we consider the tensor to scalar ratio  $r=0.001$ together with  the lower and upper bounds of speed of sound at the end of inflation $1/\sqrt{3}<c_{s_f}\leq 1 $, we find  that  the number of $e-$folds at the end of inflationary epoch  $N_f$ is in the range $0.62\leq N_f<0.84$. In this sense, we find that the number of $e-$folds at the end of inflation $N_f$ becomes 
$N_f\sim \mathcal{O}(1)$.

In relation to the consistency relation i.e., $r=r(n_s)$, combining Eqs.(\ref{nn}) and (\ref{r(n)}), we find that this relation can be written as 
\begin{equation}
\label{r(ns)}
    r(n_s)\simeq \frac{2 \alpha\,\tilde{c}_{s_f}(1-n_s)^2}{C}\left[1-\frac{2^{\beta-1}\,\alpha}{(1-\beta)C}(1-n_s)^{1-\beta}\right]^{-1}.
\end{equation}

\begin{figure}
 	\centering
\includegraphics[width=0.6\textwidth]{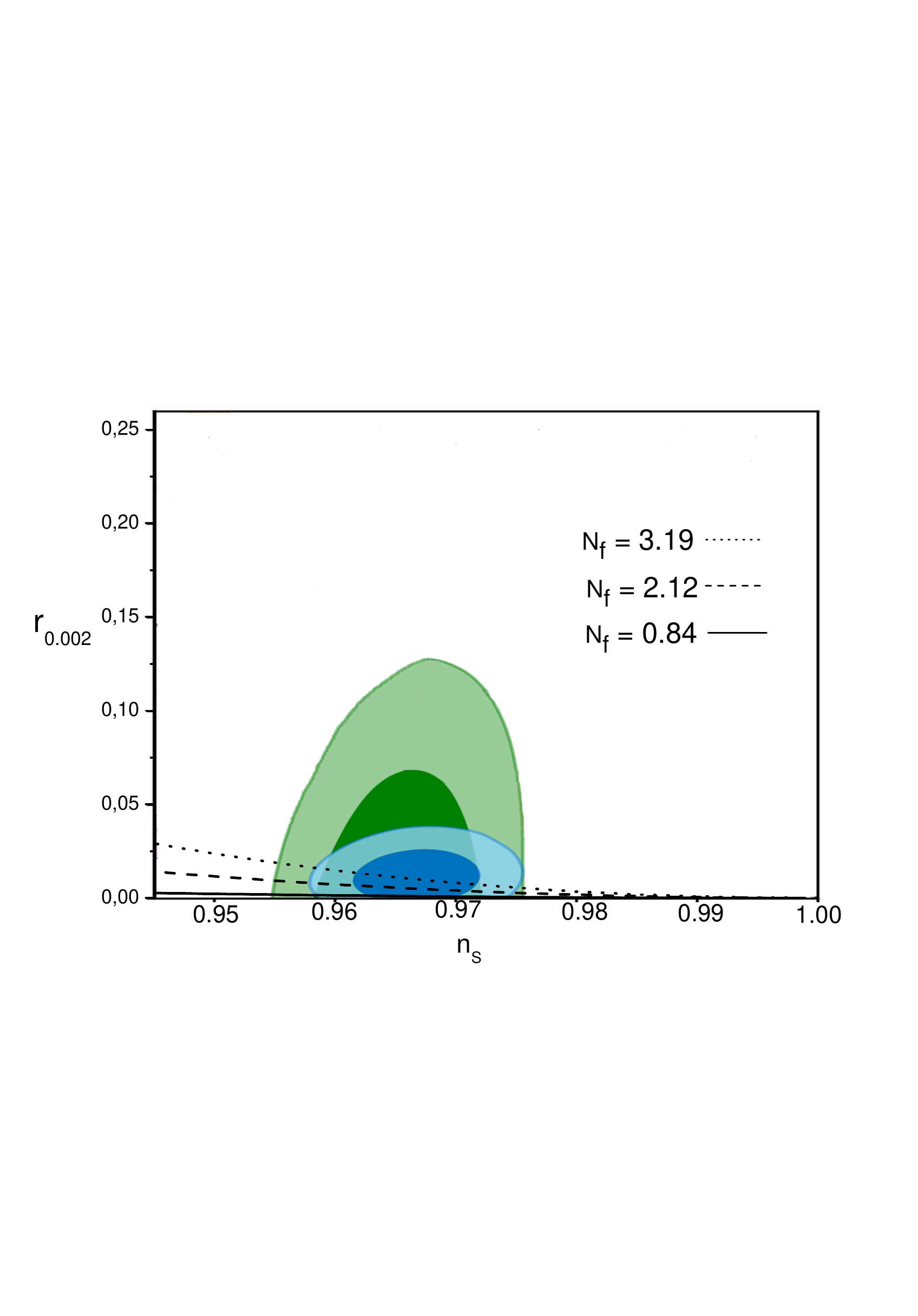}
{\vspace{-4 cm}
\caption{\it{ {The figure shows the consistency relation $r=r(n_s)$ for different 
 values of the number of $e-$folds at the end of the inflationary phase $N_f$, for the case in which we have fixed  the speed sound at the end of inflation  $c_{s_{f}}=0.6$. In this panel, the  dotted,  dashed and solid lines correspond to the pairs ($C\simeq 2.87\times 10^{9}\kappa^{2}$,\,$\alpha\simeq 2.48\times 10^{10}\kappa^{2}$) for the case in which $N_f=3.19$,    ($C\simeq 5.94\times 10^{9}\kappa^{2}$,$\,\alpha\simeq 2.42\times 10^{10}\kappa^{2}$) when $N_f=2.12$ and ($C\simeq 3.05\times 10^{10}\kappa^{2}$,\,$\alpha\simeq 2.25\times 10^{10}\kappa^{2}$) for the case in which $N_f=0.84$.
 Here we have considered the  two-dimensional marginalized joint confidence contours for the plane $(n_s,r)$ at 68\% and 95\% levels of confidence from  BICEP/Keck results \cite{BICEP:2021xfz}. \label{fig4}} }}
}	
\end{figure}

In figure \ref{fig4} we show the evolution of the tensor to scalar ratio $r$ on the scalar spectral index $n_s$ (consistency relation) for different values of the number of $e-$folds at the end of inflation $N_f$,  from the latest BICEP/Keck data \cite{BICEP:2021xfz}. For this plot we have fixed the value of the speed of sound at the end of the inflationary phase $c_{s_{f}}=0.6$. In this figure, the dotted, dashed  and solid lines correspond to the pairs ($C\simeq 2.87\times 10^{9}\kappa^{2}$,$\alpha\simeq 2.48\times 10^{10}\kappa^{2}$),  ($C\simeq 5.94\times 10^{9}\kappa^{2}$,$\alpha\simeq 2.42\times 10^{10}\kappa^{2}$) and  ($C\simeq 3.05\times 10^{10}\kappa^{2}$,\,$\alpha\simeq 2.25\times 10^{10}\kappa^{2}$), respectively.
Besides, considering  Ref.\cite{BICEP:2021xfz}, two-dimensional marginalized constraints on the tensor to scalar ratio and the scalar spectral index  fixed at $k_0=0.002$Mpc$^{-1}$. The latest BICEP/Keck results places stronger limits on the  ratio $r$ shown in blue (at 68\%  blue region and 95\%  light blue region levels of confidence) and the green color corresponds to 
 the two-dimensional marginalized constraints  found in Ref.\cite{Planck:2018vyg}.  
 
 We noted from Fig.(\ref{fig4}) that the reconstruction of our model  is well supported by the  latest BICEP/Keck  data when the speed of sound at the end of inflation $c_{s_{f}}=0.6$ and the number of $e-$folds at the end of inflation $N_f\sim \mathcal{O}(1)$. Similarly, the model  is well supported by the  Planck   data (plane $r=r(n_s)$) when the speed of sound at the end of inflation $c_{s_{f}}$
 lies in the range $1/\sqrt{3}<c_{s_{f}}\leq1$ and the number of $e-$folds at the  end of inflation $N_f\sim \mathcal{O}(1)$ (figure not shown).

\begin{table}[H]
\centering
\begin{tabular}{| c | c | c |c |c |} 
\hline
\hspace{5mm}$c_{s_f}$\hspace{5mm} & \hspace{5mm}$C\,[\kappa^2]$\hspace{5mm} & \hspace{5mm}$\alpha\,[\kappa^2]$ \hspace{5mm} & \hspace{5mm}$N_f$ \hspace{5mm} & \hspace{5mm}$\beta$ \hspace{5mm}  \\ \hline
$0.6$ & $ 3.030\times 10^{10} < C\leq$3.050$\times 10^{10}$  & $2.22\times10^{10}\leq \alpha<$2.25$\times10^{10}$ & $0.82\leq N_f<$0.84 & $-0.12\leq \beta <0.0090$\\
0.8 & $3.027 \times 10^{10}< C\leq$3.048$\times 10^{10}$  & 1.67$\times10^{10}$ $\leq \alpha<$ 1.72$\times10^{10}$& $0.70\leq N_f< 0.71$ & $-0.050\leq \beta <0.074$\\
0.99  & $3.025\times 10^{10}< C\leq$ 3.047$\times 10^{10}$ & $ 1.36\times 10^{10}\leq \alpha<$ 1.42$\times10^{10}$& $0.62\leq N_f<0.63$ & $-0.0022\leq \beta <0.12$\\ \hline
\end{tabular} 
  \caption{Summary of the allowed range 
  for the integration constants $C$ and $\alpha$ associated to the reconstructed  model  together with the range for the  parameters $N_f$ and $\beta$, when we fix different values of    the speed of sound at the end of inflation $c_{s_f}$.} \label{t1}
\end{table}

The resulting allowed values for the integration constants $C$, $\alpha$, $N_f$ and $\beta$ for the different values of the speed of sound at the end of the inflationary epoch are shown in Table \ref{t1}. Here we have considered the values $N_k=60$, $A_{s_k}=2.2\times 10^{-9}$ and $r_k=0.001$. The Table I shows that the integration constants $C$ and $\alpha$ become similar and the order of $\mathcal{O}(10^{10})\kappa^2$. Also, we observer from Table I that the number of $e-$folds at the end of the inflationary epoch $N_f\sim \mathcal{O}(1)$ and the parameter $\beta$ associated to the power of the speed of sound is close to zero. 
Besides from the Table I, we
note that the constraints on the parameters $C$, $\alpha$ and $N_f$ do not depend strongly on the value of the speed of sound at the end of inflationary epoch, since the values of these parameters are similar for all the values of $c_{s_f}$. Additionally, 
as we note from this table,  the range associated to the parameters;  $C$, $\alpha$ and $N_f$  is very narrow. However, as  we notice  the range for the parameter $\beta$ is a little larger, when we consider  different values of the parameters $C$, $\alpha$, $N_f$ and $c_{s_f}$.
Also, we note that the parameter $\beta$ can take negative values  (lower bound) depending on the values  of the speed of sound at the end of inflationary epoch. 

\section{Reheating}\label{Reheating}

In this section we will study the reheating epoch considering the background reconstructed variables  found in the previous section. In this context, we will utilize    the potential $V$ and the coupling function $g$ as a function  of the scalar field given by  Eqs.(\ref{V(phi)}) and (\ref{g(phi)}), respectively.

In the framework of a non canonical theory, we can assume that the ratio between the physical scales cross the horizon 
 during the inflationary epoch $c_sk=a_k H_k$ and the actual wave number $k_0$ related with the actual Hubble scale $H_0$ through $c_{s_0}k_0=k_0=a_0H_0$, becomes 
\begin{equation}
\label{koverk}
	\frac{c_sk}{k_0}=\frac{a_kH_k}{a_0H_0}=\left(\frac{a_k}{a_f}\right)\left(\frac{a_f}{a_\text{reh}}\right)\left(\frac{a_\text{reh}}{a_\text{eq}}\right)\left(\frac{a_\text{eq}H_\text{eq}}{a_0H_0}\right)\left(\frac{H_k}{H_\text{eq}}\right),
\end{equation}
where we have assumed that at the present the speed of sound $c_{s_0}=1$\cite{Liddle:2003as}. Also, the notation  
  ``$f$'' corresponds  to the end of inflation, ``reh'' denotes to the reheating and ``eq'' is the radiation-matter equality. Besides, we recalled that the quantities with the notation $k$  are evaluated at horizon exit during the inflationary epoch, e.g., $N\,|_{c_sk=a_kH_k}=N_k$. 
  
  Additionally, we can express the number of $e$-folds $N$ in each epoch, in terms of the scale factor $a$, as follows 
	$N_k=\ln\left[\frac{a_f}{a_k}\right]$,
	$N_\text{reh}=\ln\left[\frac{a_\text{reh}}{a_f}\right],$
 and $N_{RD}=\ln\left[\frac{a_\text{eq}}{a_\text{reh}}\right]$,
in which $N_\text{reh}$ is the number of $e$-folds during reheating era and $N_\text{RD}$ corresponds to the number of $e$-folds in the radiation dominance.

Thus, considering the different $e-$folds and   Eq.(\ref{koverk}) we get
\begin{equation}
\label{NewNreh}
	\text{ln}\left(\frac{c_sk}{a_0H_0}\right)=-N_k-N_\text{reh}-N_\text{RD}+\text{ln}\left(\frac{a_\text{eq}H_\text{eq}}{a_0H_0}\right)+\text{ln}\left(\frac{H_k}{H_\text{eq}}\right).
\end{equation}

Furthermore, we can write that   the ratio between the energy density at the end of inflationary epoch $\rho_f$ and the energy density at the end of the reheating regime $\rho_\text{reh}$ can be  associated from  an Equation of State (EoS) parameter  $\omega_\text{reh}$ related  to the reheating regime. In this form,    the ratio $\rho_\text{reh}/\rho_f$ yields
\begin{equation}
	\frac{\rho_\text{reh}}{\rho_f}=e^{-3N_\text{reh}(1+\omega_\text{reh})},
\end{equation}
in which we have assumed that during the reheating scenario the energy density decays as a function  of the scale factor  as 
 $\rho\propto a^{-3(1+\omega_\text{reh})}$, with  $\omega_\text{reh}=$ constant.

In relation to the energy density at the end of inflationary epoch  $\rho_f$ from Eq.(\ref{p1}), we have that
\begin{equation}
\label{rhoend}
	\rho_f=X_f+3g_{f}X_f^2+V_f=\left(1+\lambda\right)V_f,
\end{equation}
where the parameter $\lambda=(X_f+3g_{f}X_f^2)/V_f$. Besides, considering  the definition of the slow-roll parameter given by Eq.(\ref{s}), we find that  the parameter $\epsilon$ at the end of the inflationary stage becomes
\begin{equation}
\label{epsilon_end}
    \epsilon_f =3\frac{X_f+2g_fX_f^2}{X_f+3g_fX_f^2+V_f}.
\end{equation}
As before considering  that the end of inflation takes place 
when $\epsilon_f =1$ (or equivalently $\ddot{a}=0$), and then replacing the value of $X_f$ into the parameter $\lambda$ we get 
\begin{equation}
\lambda=1+\frac{1}{3g_fV_f}\left(1\pm\sqrt{1+3g_fV_f}\right).
\label{Lad}
\end{equation}

Thus, from Eq.(\ref{rhoend}) we find that the energy density at the end of inflation can be written as
\begin{equation}
    \rho_f=\left[\frac{3}{2}+\frac{1}{3g_fV_f}\left(1+\frac{3g_fV_f}{2}\pm\sqrt{1+3g_fV_f}\right)\right]V_f.
\label{rr1}
\end{equation}
In the following we will consider the negative signs of Eqs.(\ref{Lad}) and (\ref{rr1}), in order to recover the expressions of $\lambda$ and $\rho_f$ in the limit of a  canonical theory in which the quantity 
$g_f\rightarrow 0$. Thus, in this limit the Eqs.(\ref{Lad}) and (\ref{rr1}) with the negative signs  are reduced to $\lambda=1/2$ and  $\rho_f=3/2V_f$, respectively    
\cite{Munoz:2014eqa,Dai:2014jja,Cook:2015vqa}.

In order to determine the reheating temperature $T_\text{reh}$,  we can assume 
the entropy conservation, where the reheating entropy is preserved in the CMB together with the  neutrino background at the present \cite{Dai:2014jja}. In this sense, from the 
entropy conservation we can write \cite{Dai:2014jja} 
\begin{equation}
	g_{\text{s,reh}}a^3_{\text{reh}}T^3_{\text{reh}}=a_0^3\left(2T_0^3+\frac{21}{4}T_{\nu 0}^3\right),
\end{equation}
in which the parameter $g_{\text{s,reh}}$ denotes  the effective number of relativistic degrees of freedom for entropy at reheating, $T_0\simeq2.7$K is the current CMB temperature and the temperature $T_{\nu 0}$ corresponds to the present neutrino temperature. However, we know that 
 the relation between the neutrino temperature $T_{\nu 0}$ and $T_0$ becomes 
 $T_{\nu 0}=\left(\frac{4}{11}\right)^{1/3}T_0$ \cite{Dai:2014jja}, and then we can associate the scale factors during the reheating era and at the present epoch through 
	$a_{\text{reh}}/a_0=\left(\frac{43}{11g_{\text{s,reh}}}\right)^{1/3}T_0/T_{\text{reh}}.$
 
 Additionally, we can assume  that the energy density at the end of reheating $\rho_{\text{reh}}$ is equal  to the hot radiation and then this density yields 
\begin{equation}
\label{Treh1}
	\rho_{\text{reh}}=\frac{\pi^2}{30}g_{\star\text{,reh}}T^4_{\text{reh}}\,.
\end{equation}
Here the quantity $g_{\star\text{,reh}}$ corresponds to  the effective number of relativistic degrees of freedom at the end of reheating epoch. 

Thus, considering   the above relations,  we obtain that the reheating temperature $T_\text{reh}$ in terms of the parameters $V_f$, $g_f$, $\omega_\text{reh}$ and the number $N_\text{reh}$ results
\begin{equation}
\label{Treh}
	T_\text{reh}=\left(\frac{3}{10\pi^2}\right)^{1/4}\left[\frac{3}{2}V_f+\frac{1}{3g_f}\left(1+\frac{3g_fV_f}{2}-\sqrt{1+3g_fV_f}\right)\right]^{1/4}\text{exp}\left[-\frac{3}{4}(1+\omega_\text{reh})N_\text{reh}\right].
\end{equation}
Here the number of $e$-folds during the reheating scenario is given by 
\begin{equation}
\label{Nreh}
N_\text{reh} =\frac{4}{1-3\omega_\text{reh}}\left[- N_k-\text{ln}\left(\frac{ c_sk}{a_0T_0}\right)  -\frac{1}{3}\text{ln} \left(\frac{11g_\text{s,reh}}{43}\right)
	-\frac{1}{4}\text{ln}\left(\frac{30 \kappa^2\rho_f}{g_{\star\text{,reh}}\pi^2}\right)+\frac{1}{2}\text{ln}\left(\frac{\pi^2 r A_s}{2}\right)\right].
\end{equation}
In order to find the number $N_\text{reh}$ given by Eq.(\ref{Nreh}), we have used  that the Hubble parameter $H_k$ satisfies the relation 
between the tensor to scalar ratio $r$ and the scalar power spectrum  $A_s$ through  
$H_k=\sqrt{\frac{\pi^2}{2\kappa}rA_s}$. Here we mention that this relation for the Hubble parameter $H_k$ is still valid in the framework of a non canonical theory.

Additionally, we can rewrite Eq. (\ref{Treh}) and (\ref{Nreh}) in function of spectral index $n_s$. 
 Thus, we find that the reheating temperature and the number of $e-$folds during the reheating period as a function of the scalar spectral index $n_s$ become
\begin{equation}
\label{Trehns}
	T_\text{reh}=\left[\frac{3\rho_f(n_s)}{10\pi^2}\right]^{1/4}\text{exp}\left[-\frac{3}{4}(1+\omega_\text{reh})N_\text{reh}(n_s)\right],
\end{equation}
and 
\begin{equation}
\label{Nrehns}
N_\text{reh} =\frac{4}{1-3\omega_\text{reh}}\left[\frac{2}{(n_s-1)}-\text{ln}\left(\frac{ c_sk}{a_0T_0}\right)  -\frac{1}{3}\text{ln} \left(\frac{11g_\text{s,reh}}{43}\right)
	-\frac{1}{4}\text{ln}\left(\frac{30 \kappa^2\rho_f(n_s)}{g_{\star\text{,reh}}\pi^2}\right)+\frac{1}{2}\text{ln}\left(\frac{\pi^2 r(n_s)A_s(n_s)}{2}\right)\right],
\end{equation}
respectively.
Here the quantity $\rho_f(n_s)$ is defined as
$$
\rho_f(n_s)=\left[2+z(n_s)^{-1}\left(1-\sqrt{1+z(n_s)}\,\right)\right]\left[\frac{1}{C}-\frac{\alpha N_f^{-y(n_s)}}{y(n_s)C^2}\right],
$$
where the function $z(n_s)$ and $y(n_s)$ are given by 
\begin{eqnarray}
z(n_s)&=&\frac{18C}{\alpha}\frac{c_{s_f}^2\left(1-c_{s_f}^2\right)}{(3c_{s_f}^2-1)^2}\left[N_f^{1+y(n_s)}-\frac{\alpha N_f}{y(n_s)C}\right],\,\,\,\,\,\,\,\,\,\,\,\,\,\,\,\,\text{and}\\
y(n_s)&=&\frac{\text{ProductLog}[\alpha_2\ln(2(1-n_s)^{-1}N_f^{-1})]}{\ln[2(1-n_s)^{-1}N_f^{-1}]}.
\end{eqnarray}

Besides, the tensor to scalar ratio in function of the scalar spectral index (consistency relation) $r=r(n_s)$ is given by Eq.(\ref{r(ns)}) and the power spectrum $A_s(n_s)$ from Eq.(\ref{AS}) becomes
\begin{equation}
A_s(n_s)=\frac{\kappa^2}{3\alpha \pi^2 \tilde{c}_{s_f}(1-n_s)^2}\left[1-\frac{2^\beta\alpha (1-n_s)^{1-\beta}}{(1-\beta)C}\right].
\label{As4}
\end{equation}
Let us now consider the fact that the energy density of reheating has to be less than the energy density at the end of the inflationary epoch. In this way, we can write  the following equation \cite{Chiba:2015zpa}
\begin{equation}
    \frac{\pi^2}{30}g_{\star\text{,reh}}T^4_\text{reh}<\rho_f=\frac{\rho_f}{V}\frac{3\pi^2 r A_s}{2\kappa^2},
\end{equation}
where we have considered that the effective potential $V=3\pi^2rA_s/(2\kappa^2)$.

Thus, if we consider the upper limit of this inequality in which $(\pi^2/30)g_{\star,\text{reh}}T_\text{reh}^{\text{c}^4}=\rho_f$, we can obtain the so-called critical temperature $T_\text{reh}^\text{c}$ of the model and this temperature becomes
\begin{equation}
    T_\text{reh}^\text{\,c}\simeq \left[\frac{3}{2}+\frac{1}{3g_fV_f}\left(1+\frac{3g_fV_f}{2}-\sqrt{1+3g_fV_f}\right)\right]^{1/4}\,\left[1-\frac{(1-n_s)^{2y(n_s)}}{2^{2y(n_s)}y(n_s)^2}\left(\frac{\alpha}{C}\right)^2\right]^{1/4}\,\left[\frac{30\,V_f }{g_{\star\text{,reh}}\pi^2}\right]^{1/4},
\end{equation}
where we have used   Eqs.(\ref{V(N)aprox}), (\ref{r(ns)}) and (\ref{As4}), respectively. In particular considering the special case  in which $c_{s_{f}}=0.8$ together with the values of the parameters of the Table I,  we get
\begin{equation}
    T_\text{reh}^\text{\,c}\simeq 1.94\times 10^{15} \text{GeV}\left[1-\frac{(1-n_s)^{2y(n_s)}}{2^{2y(n_s)}y(n_s)^2}\left(\frac{\alpha}{C}\right)^2\right]^{1/4},
\end{equation}
where we have assumed that $g_{\star\text{,reh}}=106.75$. 

\begin{figure}[h]
\includegraphics[width=0.6\linewidth]{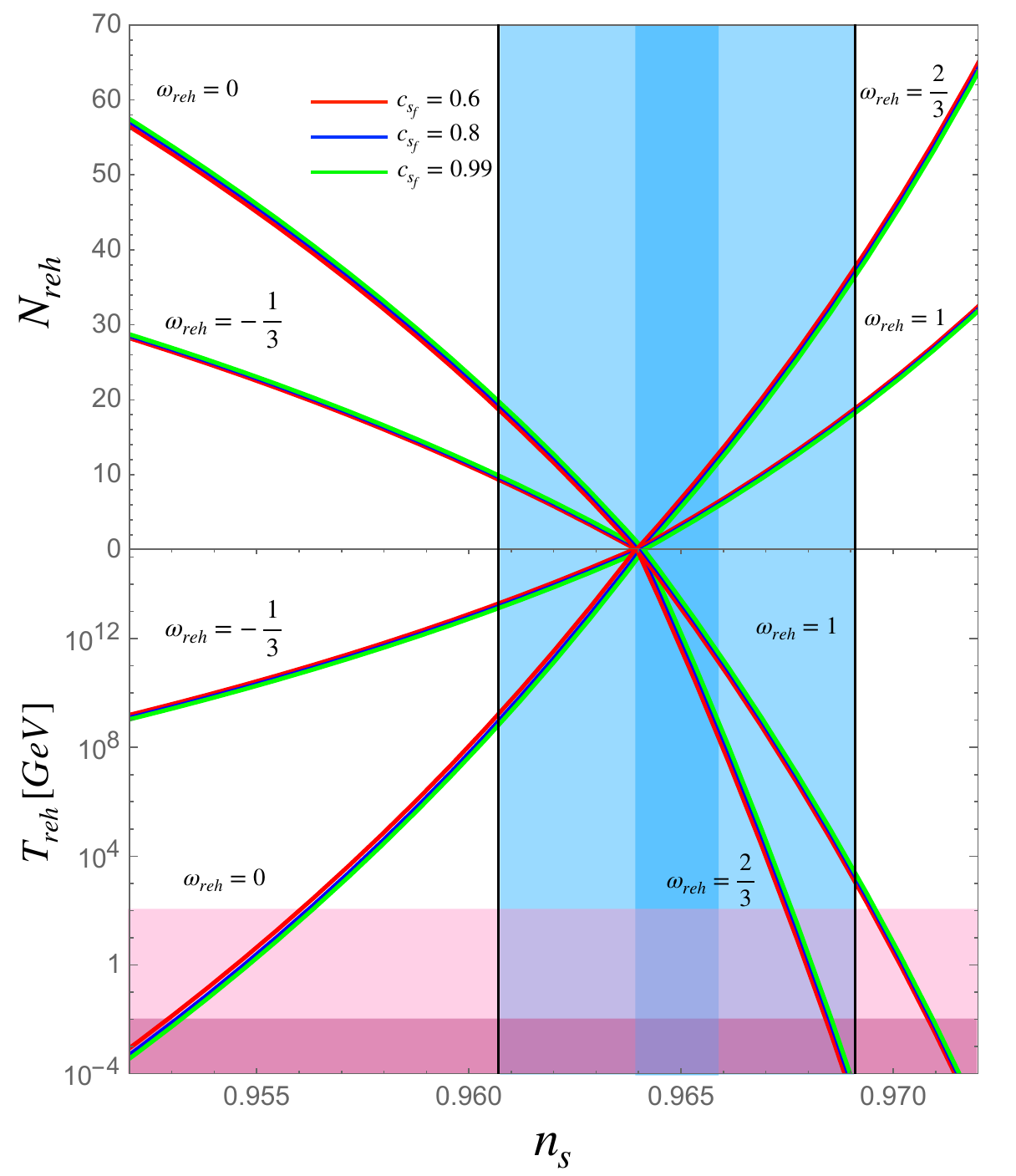}
  \caption{ The figure shows the number of $e-$ folds during the reheating scenario $N_\text{reh}$ (upper panel) and the reheating temperature $T_\text{reh}$ (lower panel)  versus the scalar spectral index $n_s$ for different speed of sound at the end of inflationary epoch $c_{s_f}$. For the value $c_{s_f}=0.6$ corresponds to the red line, $c_{s_f}=0.8$ to blue line and $c_{s_f}=0.99$ to green line, respectively. 
 Besides,  in both panels we have assumed different values of the EoS parameter $\omega_\text{reh}$; $\omega_\text{reh}=-1/3,0,2/3$ and $\omega_\text{reh}=1$, respectively. Also we have used  the parameters shown in Table I (upper bounds). }
\label{T_N_reh}
\end{figure}

In Fig.\ref{T_N_reh} we show 
the evolution   the number of $e-$folds during reheating $N_\text{reh}$ (upper panel) and  the reheating temperature $T_\text{reh}$ (lower panel) in terms of the scalar spectral index $n_s$. Here 
different values have been used for the EoS parameter during the reheating period; $\omega_\text{reh}=\{-1/3, 0,2/3,1\}$ together with some values of the speed of sound at the end of the inflationary era; $c_{s_{f}}=\{0.6,0.8,0.99\}$. In this context, for each value of the EoS parameter  $\omega_\text{reh}$, we have graphed   three 
 propagation speed of sound at the end of inflation $c_{s_f}$.

The different regions of the  Fig.\ref{T_N_reh} correspond to; the light blue shaded region
denotes  
the maximum and minimum limits of the scalar spectral index $n_s=0.9649\pm 0.0042$ from  Planck data  at 1$ \sigma$ bounds. Besides,  the blue shaded corresponds to a projected sensitivity from the central value  of the scalar spectral index at $n_s=0.9649$  of $\pm 10^{-3}$ in the test studied in Ref. \cite{-3}. Further, 
the pink shaded region denotes  the electroweak scale in which the temperature is approximately  $T_\text{EW}\sim100\text{GeV}$ and the purple shaded region with temperatures below   10 MeV,  and this region is rejected  by primordial nucleosynthesis. 
From  Fig.\ref{T_N_reh},   the point in which all the lines converge corresponds to the instantaneous reheating where the number of $e-$folds $N_\text{reh}\rightarrow 0$. Here we note that the EoS parameter  $\omega_\text{reh}$ does not play an important role, since all lines related to the EoS parameter $\omega_\text{reh}$ converge to the same point. Additionally, we note from Fig.\ref{T_N_reh} that the speed of sound at the end of the inflationary epoch $c_{s_f}$ does not play a significant role in the different parameters related to the reheating era, since all lines associated to a certain value of EoS parameter $\omega_\text{reh}$ are similar.

From Fig.\ref{T_N_reh}, we note that  the reconstructed model agrees with Planck's 1$\sigma$ bounds on the spectral index $n_s$ for the distinct  values of the EoS parameters  $\omega_\text{reh}$, when we consider high reheating temperatures $T_\text{reh}>10^{13}$ GeV. In particular for the case in which the  EoS parameter $\omega_\text{reh}>0$, we observe that this compatibility with the Planck-data occurs even at  low  reheating temperature. Besides, we note that any reheating temperature between the primordial nucleosynthesis  bound and the instantaneous reheating value is permitted inside the Planck's 1$\sigma$ bound regardless of the values of EoS parameter   excepting the specific value $\omega_\text{reh}=-1/3$, since this values of $\omega_\text{reh}$ presents high temperature. 
Also, we note that the reconstructed model predicts a small number of $e-$ folds during the reheating epoch ($N_\text{reh}<40$) in the range of Planck-data on the scalar spectral index.

\section{Conclusions and Remarks}\label{Conc}

In this article we have first studied the reconstruction of the background variables in the framework of a 
non-canonical theory,   assuming the parametrization  on the  cosmological parameters which are; the scalar spectral index $n_s=n_s(N)$ and the speed of sound $c_s=c_s(N)$, in terms of     the number of $e$-folds $N$. 
In the context of  the slow roll approximation, we have obtained  a general formalism of reconstruction for the background variables; the effective potential and coupling function as a function of $N$. In this general analysis, we have obtained from the parametrization of the scalar spectral index $n_s(N)$ and the speed of sound $c_s(N)$ different integral   relationships  for the effective potential and the coupling function as a function  of the number of $e$-folds $N$. 

As a specific example  to obtain the reconstructions of the scalar potential and the coupling function  in terms of the scalar field, we have considered the simplest parametrizations for both the scalar spectral index given by $n_s=1-2/N$ and the speed of sound $c_s\propto N^{-\beta}$. In fact, we have used our general expressions  given by Eqs.(\ref{VV5a}), (\ref{er1}) and (\ref{Rr2}), in order to find $V(\phi)$ and $g(\phi)$. Thus, from the  relation between the scalar field and the number of $e$-folds $N$, we have obtained the reconstruction of  
 the effective potential and the coupling function  in terms of the  scalar field, i.e., $V(\phi)$ and $g(\phi)$.
In this respect, Figs.(\ref{Fig1}) and (\ref{Fig2}) show the reconstruction of the background variables in terms of the scalar field from the simplest parametrizations $n_s(N)$ and $c_s(N)$, as well  the relation between the number  of $e-$folds $N$ and the scalar field $\phi$, i.e., $N(\phi)$ (see upper panel of Fig.{\ref{Fig1}}). Additionally, we have found the consistency relation $r=r(n_s)$ given by Eq.(\ref{r(ns)}), in which we have considered different number of $e-$folds  at the end of the inflationary scenario. Here we have obtained that the reconstruction of our model is well supported by Planck data when the number of $e-$ folds at the end of inflation 
$N_f\sim \mathcal{O}(1)$, see Fig(\ref{fig4}). Besides,  we have found the parameter-space for the different integration constants obtained from the reconstruction mechanism used, as well the constraints on the number of $e-$folds $N_f$ and the power $\beta$ assuming a specific value of the speed of sound at the end of the inflation. These results are shown in Table I. From this table, we have noted that the integration constants $C$ and $\alpha$ are similar and the constraints on these constants do not depend strongly on the value of sound speed $c_{s_f}$. In the same form, we have found that the number of $e-$ folds at the end of inflation 
$N_f\sim \mathcal{O}(1)$ and it  does not depend strongly on the value of sound speed. However, we have found that  the power $\beta$ related to the ansatz of the  speed of sound  presents a small variation between $-0.1<\beta<0.1$
for the different values 
 of sound speed $c_{s_f}$ studied.

Regarding the analysis during the reheating era  in the framework of a non-canonical theory, we have obtained that it is feasible  to describe this epoch in terms of the different reheating parameters  as;   the reheating temperature $T_\text{reh}$, number of $e-$folds associated to the reheating $N_\text{reh}$ and  the EoS parameter $\omega_\text{reh}$. Thus,
  we have found the fundamental reheating parameters given by Eqs.(\ref{Trehns}) and (\ref{Nrehns}), which are temperature and the number of $e-$folds during the reheating epoch in terms  of parameters associated to the model and these are; the sound of speed, the EoS parameter $\omega_\text{reh}$ together with 
 the observational parameters such as; $A_s$, $n_s$,  and $r$. In relation to this point, from Fig(\ref{T_N_reh}) we have found that the number of $e-$folds together with the temperature  during the reheating epoch do not depend strongly of the value of the speed of sound at the end of inflationary scenario. Here the red, blue and green lines that indicate the different values of the speed of sound $c_{s_{f}}$ are similar for a given value of the EoS parameter  $\omega_\text{reh}$.

In relation to the Planck data  at 1$\sigma$ bound on the scalar spectral index $n_s$, we have obtained that for the situations in which the EoS parameter  $\omega_\text{reh}\leq 0$, the reconstructed model shows  higher values associated to the reheating temperatures, in particular for the case in which the EoS parameter $\omega_\text{reh}=-1/3$, we have found that  $T_\text{reh}\gtrsim 10^{12}$GeV. 
Additionally, we have found that the duration of the reheating epoch distinguished   by   the number of $e$-folds $N_\text{reh}$ turns out to be  small that  $N_\text{reh}\lesssim 35$ for the different values of the EoS parameter $\omega_\text{reh}$
 from the constraints given by Planck data  at 1$\sigma$ bound on $n_s$, see Fig.(\ref{T_N_reh}). 

Finally   in this work, we have not addressed the 
reheating study in  numerical form assuming   an EoS parameter $\omega_\text{reh}$
that depends on time,  to analyze the reheating parameters; $T_\text{reh}$ and $N_\text{reh}$. In this
point, we hope to return to this study in the near future.

\section{acknowledgments}
C. R. thanks to the Vicerrector\'{\i}a de Investigaci\'on Creaci\'on e Innovaci\'on, Pontificia Universidad Cat\'olica de Valpara\'{\i}so,  Postgraduate Scholarship PUCV 2024.
\\
\\
\\
\\
\\

\end{document}